\documentclass[superscriptaddress,onecolumn,pre,nofootinbib]{revtex4}

\usepackage{amsmath}
\usepackage{graphicx,color}

\newcommand{\be}{\begin{equation}}
\newcommand{\ee}{\end{equation}}
\newcommand{\bea}{\begin{eqnarray}}
\newcommand{\eea}{\end{eqnarray}}

\begin{document}
\sloppy

%-title page-%

\title{Kinetic theory of spatially homogeneous systems with long-range interactions: \\
III. Application to power-law potentials, plasmas, stellar systems, \\
and to the HMF model}

\author{Pierre-Henri Chavanis}
%\email{chavanis@irsamc.ups-tlse.fr}
\affiliation{Laboratoire de Physique Th\'eorique (IRSAMC), CNRS and UPS, Universit\'e de Toulouse, F-31062 Toulouse, France}

\begin{abstract}

We apply the general results of the kinetic theory of systems with long-range interactions to particular systems of physical interest. We consider repulsive and attractive power-law potentials of interaction $r^{-\gamma}$ with $\gamma<d$ in a space of dimension $d$. For $\gamma>\gamma_c\equiv (d-1)/2$, strong collisions must be taken into account and the evolution of the system is governed by the Boltzmann equation or by a modified Landau equation; for $\gamma<\gamma_c$, strong collisions are negligible and the evolution of the system is governed by the Lenard-Balescu equation. In the marginal case $\gamma=\gamma_c$, we can use the Landau equation (with appropriately justified cut-offs) as a relevant approximation of the Boltzmann and Lenard-Balescu equations. The divergence at small scales that appears in the original Landau equation is regularized by the effect of strong collisions. In the case of repulsive interactions with a neutralizing background (e.g. plasmas), the divergence at large scales that appears in the original Landau equation is regularized by collective effects accounting for Debye shielding. In the case of attractive interactions (e.g. gravity), it is regularized by the spatial inhomogeneity of the system and its finite extent. We provide explicit analytical expressions of the diffusion and friction coefficients,  and of the relaxation time, depending on the value of the exponent $\gamma$ and on the dimension of space $d$. We treat in a unified framework the case of Coulombian plasmas and stellar systems in various dimensions of space, and the case of the attractive and repulsive HMF models.

\end{abstract}

\maketitle

\section{Introduction}
\label{sec_introduction}

The dynamics and thermodynamics of systems with long-range interactions is currently a topic of active research in physics \cite{houches,assise,oxford,cdr,proceedingdenmark,bgm}. A system is said to be long-ranged if the potential of interaction decays at large distances as $r^{-\gamma}$ with $\gamma<d$. In that case, the potential energy $\int_0^{+\infty} u(r) r^{d-1}\, dr$ diverges at large scales implying that all the particles interact with each other and that the system displays a collective behavior. For such systems, the mean field approximation becomes exact
in a proper thermodynamic limit $N\rightarrow +\infty$ \cite{cdr}.
In \cite{paper1,paper2} (Papers I and II), we have developed a general framework to tackle the kinetic theory of systems with long-range interactions and we have presented the basic kinetic equations (Vlasov, Landau, Lenard-Balescu, Fokker-Planck). These equations were initially introduced in  plasma physics \cite{ichimaru,pitaevskii,nicholson,balescubook} and stellar dynamics \cite{spitzerbook,bt,hut} but they are actually valid for a much larger class of systems
with long-range interactions. In this paper, we consider specific applications of this general formalism.  We provide explicit analytical expressions of the diffusion and friction coefficients, and of the relaxation time, for plasmas and stellar systems in various dimensions of space, and for the attractive
and repulsive Hamiltonian Mean Field (HMF) models. Our approach provides a unified framework to treat these different systems. In this paper, we restrict ourselves to systems that are spatially homogeneous or for which a local approximation may be implemented. The case of spatially inhomogeneous systems is more complicated and must be treated with angle-action variables as in \cite{angleaction,kindetail,heyvaerts,newangleaction,aanew}.

We first consider repulsive and attractive power-law potentials of interaction $r^{-\gamma}$ with $\gamma<d$ in a space of dimension $d$, generalizing the traditional Coulombian and Newtonian potentials (corresponding to $\gamma=d-2$). For $\Lambda\gg 1$, where $\Lambda$ is the effective number
of particles\footnote{For attractive interactions, $\Lambda=n\lambda_J^d$ represents the number of particles
in the Jeans sphere, which is of the order of the total number of particles $N$. For repulsive interactions
with a neutralizing background, $\Lambda=n\lambda_D^d$ represents the number of particles in the Debye sphere.}, the evolution of the system is dominated by weak collisions and collective effects. We can therefore make a {\it weak coupling} approximation and expand the equations of the BBGKY hierarchy in
powers of $1/\Lambda$ (see Appendix \ref{sec_bbgky}). For $\Lambda\rightarrow +\infty$, we get the Vlasov equation. At the order $1/\Lambda$, the evolution of the distribution function is governed by the Lenard-Balescu equation. However, for singular potentials, the Lenard-Balescu equation may present divergences at small scales reflecting the importance of strong collisions that have been neglected in the weak coupling approximation. For power-law potentials,
there exist a critical index $\gamma_c=(d-1)/2$ (for Coulombian or Newtonian interactions, it corresponds to a critical dimension $d_c=3$). For $\gamma>\gamma_c$ (i.e. $d>3$ for Coulombian or Newtonian potentials), strong collisions must be taken into account while collective effects may be neglected. In that case, the evolution of the system is described
by the Boltzmann equation or by a modified Landau equation (see Appendix \ref{sec_landaumodif}). The relaxation time\footnote{Here, we consider the relaxation time of a test particle in a thermal bath. For $d>1$, it coincides with the relaxation time of the system as a whole. However, for spatially homogeneous systems in $d=1$, the collision term vanishes and the relaxation time of the system as a whole is larger than $\Lambda t_D$ (see discussion in Paper II).} scales as $t_R\sim \Lambda^{(d-\gamma-1)/\gamma}t_D$, where $t_D$ is the dynamical time. For $\gamma<\gamma_c$ (i.e. $d<3$ for Coulombian or Newtonian potentials), strong collisions are negligible while collective effects are important. In that case, the  Lenard-Balescu equation is rigorously valid. The relaxation time scales as $t_R\sim \Lambda t_D$. For $\gamma=\gamma_c$ (i.e. $d=3$ for Coulombian or Newtonian potentials), strong collisions and collective effects must be taken into account but their importance is weak. The Boltzmann equation is marginally valid provided that a large-scale cut-off is introduced at the Debye length. Similarly,  the Lenard-Balescu equation is marginally valid provided that a small-scale cut-off is introduced at the Landau length. In that marginal situation, we can use the Landau equation (with appropriately justified cut-offs) as a good approximation of the Boltzmann and Lenard-Balescu equations. The relaxation
time scales as $t_R\sim (\Lambda/\ln \Lambda) t_D$. These scalings
agree with those obtained by Gabrielli {\it et al.} \cite{gjm} based on an extension of the Chandrasekhar
binary collision theory. The binary collision theory is usually adapted
to systems with short-range interactions. It can be used when $\gamma>\gamma_c$ and it is marginally 
valid when $\gamma=\gamma_c$ (corresponding to 3D Coulombian or 3D
Newtonian potentials). For $\gamma\le \gamma_c$ it produces a divergence at large scales. By contrast,
the Lenard-Balescu theory is adapted to systems with long-range interactions. It can be used when 
$\gamma<\gamma_c$ and it is marginally valid when $\gamma=\gamma_c$.  For $\gamma\ge \gamma_c$ it produces 
a divergence at small scales. Therefore, these two theories are 
complementary to each other.

We discuss in detail the physical regularization of the  divergences that occur in the Landau equation. The
small-scale divergence is regularized by the effect of strong collisions. They are taken into account
in the Boltzmann equation or in the modified Landau equation. For repulsive potentials with a neutralizing background (as in plasma physics), the large-scale divergence is regularized by collective effects accounting for Debye shielding. In a plasma, since the Coulomb force between like-sign (resp. opposite-sign) charges is repulsive (resp. attractive), the sign of the polarization cloud surrounding  a test charge is opposite to that of the test charge. As a result, the interaction is screened on a distance of the order of the Debye length $\lambda_D$ \cite{dh}. Collective effects are accounted for in the Lenard-Balescu equation. This equation does not present any divergence at large scales, contrary to the Landau equation, and the Debye length appears naturally. For attractive potentials (as in stellar dynamics), the large-scale divergence is regularized by the spatial inhomogeneity of the system and its finite extent.
The Jeans length $\lambda_J$ \cite{jeansbook}, which is of the order of the system's size, represents a natural large-scale cut-off. Spatial inhomogeneity is accounted for in the Landau and Lenard-Balescu equations written in angle-action variables \cite{angleaction,kindetail,heyvaerts,newangleaction,aanew}. These equations do not present any divergence at large scales, contrary to the case where we make a local approximation. In the gravitational case, collective effects lead to a form of anti-shielding. The test star draws neighboring stars into its vicinity and these add their gravitational force to that of the test star itself. The ``bare'' gravitational force of the test star is thus augmented rather than shielded. The polarization acts to increase the effective gravitational mass of a test star. As a result, collective effects tend to increase the value of the diffusion coefficient and reduce the relaxation time of the system \cite{gilbert,weinberg,aanew}.

For 3D plasmas, the Lenard-Balescu equation is marginally valid provided that a small-scale cut-off is introduced at the Landau length in order to take into account the effect of strong collisions. The relaxation time scales as $(\Lambda/\ln\Lambda) t_D$ where $\Lambda$ is the number of electrons in the Debye sphere. The Landau equation with a small-scale cut-off at the Landau length and a large-scale cut-off at the Debye length provides an excellent approximation of the Lenard-Balescu equation.  For 2D plasmas, the Lenard-Balescu equation is rigorously valid. The relaxation time scales as $\Lambda t_D$. The Landau equation with a cut-off at the Debye length is a poor approximation of the Lenard-Balescu equation but the discrepancy  is not dramatic. In particular, the velocity dependence of the diffusion coefficient is the same, up to a numerical factor. Therefore, we may use the Landau equation by ``adapting'' the value of the large scale cut-off (it is a fraction of the Debye length).  For 1D plasmas, the Lenard-Balescu equation is rigorously valid but it trivially reduces to zero. The relaxation time of the system as a whole scales as $\Lambda^2 t_D$ while the relaxation time of a test particle in a thermal bath scales as $\Lambda t_D$. The Landau equation leads to wrong results. It predicts that the diffusion coefficient of a test particle in a thermal bath decays like a Gaussian while in reality, when collective effects are properly accounted for, it decays as $|v|^{-1}$.

For 3D stellar systems, the Lenard-Balescu equation and the  Landau equation written with angle-action variables are marginally valid provided that a small-scale cut-off is introduced at the Landau length in order to take into account the effect of strong collisions. The relaxation time scales as $(N/\ln N) t_D$ where $N$ is the number of stars in the cluster. The Landau equation based on a local approximation with a small-scale cut-off at the Landau length and a large-scale cut-off at the Jeans length is relatively accurate.  For 2D stellar systems, the Lenard-Balescu equation and the  Landau equation  written with angle-action variables are rigorously valid. The relaxation time scales as $N t_D$. The Landau equation based on a local approximation produces a linear divergence at large scales but it may be used provided that a large scale cut-off is introduced and properly ``adapted'' (it is a fraction of the Jeans length). For 1D stellar systems, the Lenard-Balescu equation and the Landau equation  written with angle-action variables are rigorously valid. The relaxation time scales as $N t_D$. The Landau equation based on a local approximation leads to wrong results. In particular, it predicts a relaxation time of the system
as a whole scaling as $N^2t_D$ while the right scaling is $Nt_D$ when spatial inhomogeneity is accounted for.

For the HMF model, explicit expressions of the diffusion and friction coefficients, and of the  relaxation time, can be obtained in the homogeneous phase. The relaxation time of the system as a whole is larger than $Nt_D$ (scaling presumably as $N^2 t_D$) while the relaxation time of a test particle in a bath scales as $Nt_D$. The diffusion coefficient depends on the temperature. For the attractive HMF model, there
is a critical point ($E_c,T_c)=(3/4,1/2)$ and the diffusion coefficient increases close to the critical point implying a decrease of the 
relaxation time. 

Some results of kinetic theory have been obtained previously in particular cases, using
different approaches, and we give the corresponding references. However, the originality of our approach is to develop a general framework to treat the kinetic theory of systems with long-range interactions (Papers I and II) and, from this framework, consider particular systems. This provides a more {\it unified} description of systems with long-range interactions.

The paper is organized as follows. In Sec. \ref{sec_pot}, we consider power-law potentials of interactions 
and introduce appropriate notations. In Sec. \ref{sec_powerlaw}, we provide an estimate of the relaxation 
time for systems interacting with power-law potentials and we discuss how it depends on the value of the 
exponent $\gamma$. In Sec. \ref{sec_plasmas} and \ref{sec_sg}, we treat the case of Coulombian plasmas 
and self-gravitating systems in $d$ dimensions. In Sec. \ref{sec_hmf}, we consider the attractive and 
repulsive HMF models. The Appendices gather additional results of kinetic theory that complete the 
main discussion of the paper.

\section{Power-law potentials of interaction}
\label{sec_pot}

In this section, we  consider power-law potentials of interaction of the form $r^{-\gamma}$ with $\gamma<d$
in a space of dimension $d$. We introduce appropriate notations that generalize those commonly used in the study of plasmas
and self-gravitating systems.

\subsection{Repulsive interactions: plasmas}
\label{sec_potrep}

The potential of interaction of a one-component Coulombian plasma with a neutralizing background in $d$-dimensions is the solution of the Poisson equation $\Delta u=-S_d(e^2/m^2)\delta({\bf r})$ where $S_d$ is the surface of a unit sphere, $-e$ is the charge of the electron, and $m$ is its mass. Its Fourier transform is $(2\pi)^d\hat{u}(k)=S_d e^2/(m^2k^{2})$.

More generally, we consider a repulsive power-law potential whose
Fourier transform may be written as $(2\pi)^d\hat{u}(k)=S_d e^2/m^2k^{2-\alpha}$. The potential of interaction in physical space is of the form  $u({\bf r})=K_{d,\alpha} (S_d e^2/m^2)/|{\bf r}|^{d-2+\alpha}$,
where $K_{d,\alpha}=\frac{1}{(2\pi)^d}\int d{\bf q}\, e^{i{\bf q}\cdot \hat{\bf r}}/q^{2-\alpha}=\pi^{-d/2} 2^{\alpha-2}\Gamma((d-2+\alpha)/2)/\Gamma((2-\alpha)/2)$ is a constant depending on $d$ 
and $\alpha$ (its precise expression is not needed since only the Fourier transform of the potential of interaction matters in the kinetic theory). The exponent of
power-law decay is $\gamma=d-2+\alpha$. The Coulombian potential corresponds to $\alpha=0$, i.e. $\gamma=d-2$. For $\alpha=2-d$ (i.e. $\gamma=0$), $u({\bf r})=K_{d}(e^2/m^2) \ln |{\bf r}|$ is the logarithmic potential.

We introduce the parameter $\beta=1/(m v_m^2)$ where $v_m$ is a typical velocity of the system (for a system at statistical equilibrium, $\beta=1/k_B T$ is
the inverse temperature). We define the Debye wavenumber $k_D$
and the  plasma pulsation $\omega_P$ by the relations $k_D^{2-\alpha}=S_d e^2\beta\rho/m$ and
$\omega_P^2=S_d\rho e^2 k_D^{\alpha}/m^2$ where $\rho=n m$ is the mass density.
They are related to each other by $\omega_P=k_D/\sqrt{\beta m}=k_D v_m$. We also introduce the
dynamical time $t_D=\omega_P^{-1}=\sqrt{\beta m}/k_D=1/(k_D v_m)=1/\sqrt{S_d\rho e^2 k_D^{\alpha}/m^2}$. It represents the typical
time needed by a particle to travel the Debye length $\lambda_D\sim k_D^{-1}$ with
the velocity $v_m\sim (\beta m)^{-1/2}$. Finally, we introduce the parameter $\Lambda=n \lambda_D^{d}$ that gives the number of electrons in the Debye
sphere. The normalized potential $\eta(k)=-(2\pi)^d\hat{u}(k)\beta m\rho$ introduced in Paper  II may be written as $\eta(k)=-(k_D/k)^{2-\alpha}$.

The number of charges in the Debye sphere scales like $\Lambda\sim n^{(2-\alpha-d)/(2-\alpha)}(T/e^2)^{d/(2-\alpha)}$. For the Coulombian potential ($\alpha=0$), we get
 $\Lambda\sim n^{(2-d)/2}T^{d/2}/e^d$. The number of charges in the Debye sphere is always an increasing function of the temperature. By contrast, its dependence with the density depends on the dimension of space. In $d=3$, $\Lambda\sim n^{-1/2}T^{3/2}/e^3$ decreases with the density. In $d=2$, $\Lambda\sim T/e^2$ does not depend on the density. In $d=1$, $\Lambda\sim n^{1/2}T^{1/2}/e$ increases with the density.

{\it Remark 1:} in the previous formulae, $e$ represent  the charge of the electron {\it only} when $\alpha=0$. However, when $\alpha\neq 0$, it is always possible to write the potential of interaction in the above form (where $e$ may be regarded as a coupling constant) so as to facilitate the connection with Coulombian plasmas when $\alpha=0$.

\subsection{Attractive interactions: self-gravitating systems}
\label{sec_potatt}

The potential of interaction of a self-gravitating system in $d$-dimensions is the solution of the Poisson equation $\Delta u=S_d G\delta({\bf r})$ where $G$ is the gravitational constant. Its Fourier
transform is $(2\pi)^d\hat{u}(k)=-S_d G/k^2$.

More generally, we consider an attractive power-law potential whose
Fourier transform may be written as $(2\pi)^d\hat{u}(k)=-S_d G/ k^{2-\alpha}$. The potential of interaction in physical space is $u({\bf
r})=K_{d,\alpha} S_d G/|{\bf r}|^{d-2+\alpha}$. The exponent of power-law decay is $\gamma=d-2+\alpha$. The Newtonian potential corresponds to $\alpha=0$, i.e. $\gamma=d-2$. For
$\alpha=2-d$ (i.e. $\gamma=0$), $u({\bf r})=K_d G \ln |{\bf r}|$ is the logarithmic
potential.

We introduce the parameter $\beta=1/(m v_m^2)$ where $v_m$ is a typical velocity of the
system (for a system at statistical equilibrium, $\beta=1/k_B T$ is
the inverse temperature).  We define the Jeans wavenumber $k_J$ and the gravitational pulsation $\omega_G$ by the relations $k_J^{2-\alpha}=S_d G\beta m \rho$ and
$\omega_G^2=S_d G \rho k_J^{\alpha}$ where $\rho=n m$ is the mass density.  They are related to each other
by $\omega_G=k_J/\sqrt{\beta m}=k_J v_m$. We also introduce the
dynamical time $t_D=\omega_G^{-1}=\sqrt{\beta m}/k_J=1/(k_J v_m)=1/\sqrt{S_d G\rho k_J^{\alpha}}$.  It represents the typical time
needed by a particle to travel the Jeans length $\lambda_J\sim k_J^{-1}$ with the velocity $v_m\sim (\beta
m)^{-1/2}$. Finally, we introduce the parameter $\Lambda=n \lambda_J^{d}$ that
gives the number of particles in the Jeans sphere. The normalized potential
$\eta(k)=-(2\pi)^d\hat{u}(k)\beta m\rho$  may be written as $\eta(k)=(k_J/k)^{2-\alpha}$.

For attractive power-law potentials in $d$ dimensions with $\alpha\neq 2-d$, the virial theorem 
reads $2K+(d-2+\alpha)W=0$ where $K$ is the kinetic energy and $W$ the potential energy (see Appendix I 
of \cite{n2}). The kinetic energy may be estimated by $K\sim N m v_m^2$ and the potential
energy by $|W|\sim N^2 m^2 G / R^{d-2+\alpha}$ where $R$ is the system's size. Since the virial theorem implies  $K\sim |W|$, we find from the foregoing expressions that $\beta GMm/R^{d-2+\alpha}\sim 1$.  When $\alpha=2-d$, the virial theorem reads $2K-GM^2/2=0$ leading to the exact result $\langle v^2\rangle=GM/2$ (the velocity dispersion can take a {\it unique} value in a steady state) \cite{n2}.  At statistical equilibrium $K=dNk_BT/2$, so we get $\beta GMm=2d$. Using $\rho\sim M/R^d$, we find in all cases that $\lambda_J\sim R$. Therefore, the Jeans length $\lambda_J$ represents the typical size $R$ of the system. Accordingly, the parameter $\Lambda=n
\lambda_J^{d}\sim N$ gives the total number of particles in the system.

{\it Remark 2:} in the previous formulae, $G$ represent  the gravitational constant {\it only} 
when $\alpha=0$. However, when $\alpha\neq 0$, it is always possible to write the potential 
of interaction in the above form (where $G$ may be regarded as a coupling constant) so as to 
facilitate the connection with self-gravitating systems when $\alpha=0$. The statistical mechanics
of systems with attractive power-law interactions has been studied in \cite{ic}.

\subsection{Characteristic lengths}
\label{sec_lengths}

There are three characteristic lengths in the problem:

$\bullet$ The length $l\sim n^{-1/d}$, where $n$ is the numerical density, gives the typical distance
between particles.

$\bullet$ The Debye length $\lambda_D=(m/S_d e^2
\beta\rho)^{1/(2-\alpha)}$ for repulsive potentials gives the effective
range of the interaction due to the screening by opposite charges,
and the Jeans length $\lambda_J=(1/S_d G\beta m\rho)^{1/(2-\alpha)}$
for attractive potentials gives the typical size $R$ of the system.
These expressions rely on a mean field approximation so they are valid
only for long-range interactions ($\gamma<d$ i.e.  $\alpha<2$). We shall denote commonly the Debye and Jeans lengths by $\lambda_s$ and define $\Lambda=n\lambda_s^d$.

$\bullet$ The Landau length $\lambda_L$ is the distance at which
binary collisions become ``strong''. This corresponds to an impact parameter yielding a
 deflexion at $90^o$. This happens when the energy of
interaction $e^2/\lambda^{d-2+\alpha}$ or
$Gm^2/\lambda^{d-2+\alpha}$ between two particles becomes
comparable to their kinetic energy $mv_m^2$. Equating these two
quantities, we get $\lambda_{L}\sim
(e^2/m v_m^2)^{1/(d-2+\alpha)}$ and $\lambda_{L}\sim
(Gm/v_m^2)^{1/(d-2+\alpha)}$. These definitions make sense only for decaying potentials ($\gamma>0$ i.e. $\alpha>d-2$). Introducing the notations defined in the preceding sections, the Landau wavenumber may be rewritten as $k_{L}\sim
\left ({n}/{k_s^{2-\alpha}}\right )^{1/(d-2+\alpha)}\sim
\Lambda^{1/(d-2+\alpha)}k_s$. For attractive
interactions, using $\lambda_J\sim R$ and $\Lambda\sim N$, we get
$\lambda_{L}\sim R/N^{1/(d-2+\alpha)}$.

Combining the previous expressions, we obtain
\begin{eqnarray}
\label{est}
\left (\frac{\lambda_s}{l}\right )^d\sim \left (\frac{\lambda_s}{\lambda_L}\right )^{d-2+\alpha}\sim \left (\frac{l}{\lambda_L}\right )^{\frac{d(d-2+\alpha)}{2-\alpha}}\sim \Lambda.
\end{eqnarray}
For $2-d<\alpha<2$ (i.e. $0<\gamma<d$) and $\Lambda\gg 1$, we find that $\lambda_L\ll l\ll \lambda_s$.

The coupling parameter $\Gamma=E_{pot}/E_{kin}$ is equal to the ratio
between the potential energy $E_{pot}\sim e^2/n^{-(d-2+\alpha)/d}$ or
$E_{pot}\sim Gm^2/n^{-(d-2+\alpha)/d}$ of two particles separated by the average
inter-particle distance $l\sim n^{-1/d}$ and the typical kinetic energy
of a particle $E_{kin}\sim mv_m^2=1/\beta$. This yields $\Gamma=e^2 n^{(d-2+\alpha)/d}/mv_m^2$ or $\Gamma=Gm n^{(d-2+\alpha)/d}/v_m^2$.  Using the previous relations, we find that
$\Gamma\sim 1/\Lambda^{(2-\alpha)/d}$. The weak coupling approximation corresponds to
$\Gamma\ll 1$. For long-range interactions ($\gamma<d$ i.e. $\alpha<2$), the conditions $\Lambda\gg 1$ and $\Gamma\ll 1$ are equivalent. We may also define the coupling parameter $g=E_{pot}^*/E_{kin}$ where $E_{pot}^*\sim e^2/\lambda_D^{d-2+\alpha}$ or $E_{pot}^*\sim G m^2/\lambda_J^{d-2+\alpha}$ is the potential energy of two particles separated by the Debye length or by the Jeans length. In that case, we find $g\sim 1/\Lambda$. The parameter $g$ is called the plasma parameter or the graininess parameter. It is equal to the reciprocal of the number of particles in the Debye or Jeans sphere.

\section{Estimate of relaxation time for systems with power-law potentials}
\label{sec_powerlaw}

The normalized Fourier transform of the power-law potentials defined previously may be written as
\begin{eqnarray}
\eta(k)=\epsilon \left (\frac{k_s}{k}\right )^{2-\alpha},
\label{pl1}
\end{eqnarray}
where $\epsilon=+1$ for attractive interactions and $\epsilon=-1$ for repulsive interactions (in the repulsive case, we assume that there exist a neutralizing background that maintains the spatial homogeneity of the system). The Coulombian potential (plasmas) corresponds to $\epsilon=-1$, $\alpha=0$ and  $k_s=k_D$ (Debye wavenumber) and the Newtonian potential (gravity) corresponds to $\epsilon=+1$, $\alpha=0$  and $k_s=k_J$ (Jeans wavenumber).

To estimate the relaxation time, we consider the Landau equation (II-5)\footnote{Here and in the following, (II-n) refers to Eq. (n) of Paper II.} in which strong collisions and
collective effects are neglected (see Appendix \ref{sec_bbgky} for more details about the approximations
made in this section). Therefore, we only consider weak collisions. We also assume that the system
is spatially homogeneous or that a local approximation may be implemented. In the thermal bath approach, the
diffusion coefficient is given by Eq. (II-45) with
\begin{equation}
\label{pl2}
{\cal D}_d=\frac{1}{2(2\pi)^{d-\frac{1}{2}}}\frac{v_m^3}{n}k_s^{2(2-\alpha)}\int_0^{+\infty} \frac{dk}{k^{4-d-2\alpha}}.
\end{equation}
The relaxation time may be estimated by $t_R^{bath}\sim v_m^2/{\cal D}_d$ (see Paper II). Defining the dynamical time by $t_D\sim \omega_s^{-1}\sim (v_m k_s)^{-1}$ and performing the change of variables $\kappa=k/k_s$, we obtain
\begin{equation}
\label{pl3}
t_{R}^{bath}\sim \frac{\Lambda}{\int_0^{+\infty} \frac{d\kappa}{\kappa^{4-d-2\alpha}}}t_D,
\end{equation}
where $\Lambda=nk_s^{-d}$. For  repulsive interactions $\Lambda= n k_D^{-d}$ represents the number of particles in the Debye sphere and for attractive interactions $\Lambda= n k_J^{-d}\sim N$ represents the total number of particles in the system. The relaxation
time scales like $\Lambda t_D$ except if the integral in Eq. (\ref{pl2}) diverges. In that case, it must be regularized at small or large scales and the regularized integral may depend on $\Lambda$. We must distinguish different cases according to whether $\alpha$ (or $\gamma$) is larger, smaller, or equal to the critical index
\begin{equation}
\label{pl4}
\alpha_c=\frac{3-d}{2}\qquad \left ({\rm or} \qquad \gamma_c=\frac{d-1}{2}\right ).
\end{equation}
For a Coulombian or a Newtonian potential ($\alpha=0$, $\gamma=d-2$), this critical index corresponds to the critical dimension $d_c=3$.

\subsection{The case $\alpha_c<\alpha<2$}
\label{sec_mls}

If $\alpha_c<\alpha<2$ (i.e. $\gamma_c<\gamma<d$ or $d>3$ for  a Coulombian or a Newtonian potential), the integral (\ref{pl2})  converges for $k\rightarrow 0$ (large scales) implying  that collective effects (and spatial inhomogeneity effects for attractive interactions) are weak. On the other hand, it diverges algebraically for $k\rightarrow +\infty$ (small scales). This divergence is regularized by taking strong collisions into account. Heuristically, we can introduce a small-scale cut-off at the Landau length  at which collisions become strong. Performing the integrals in Eqs. (\ref{pl2}) and (\ref{pl3}) with $k_{max}\sim k_{L}=\Lambda^{\frac{1}{d-2+\alpha}}k_s$, we obtain
\begin{equation}
\label{pl5}
{\cal D}_d=\frac{1}{2(2\pi)^{d-\frac{1}{2}}}\, \frac{1}{d-3+2\alpha}\, \frac{v_m^3 k_s}{\Lambda^{\frac{1-\alpha}{d-2+\alpha}}},\qquad t_R^{bath}\sim \Lambda^{\frac{1-\alpha}{d-2+\alpha}} t_D.
\end{equation}

We now develop a more precise description (see Appendix \ref{sec_bbgky}).
In the limit $\Lambda\gg 1$, the evolution of the system is dominated by weak collisions and collective effects (and spatial inhomogeneity for attractive interactions). Three-body collisions are negligible. In principle, strong collisions should also be negligible but the divergence of the diffusion coefficient (\ref{pl2}) shows that they are important at small scales. Since the divergence is strong (algebraic), the Lenard-Balescu equation is not applicable. On the other hand, collective effects and spatial inhomogeneity do not seem to be crucial since the diffusion coefficient (\ref{pl2}) converges at large scales. If we only take weak and strong collisions into account, we obtain the Boltzmann equation. This equation does not present any divergence. Since weak collisions dominate over strong collisions when $\Lambda\gg 1$, we can expand the Boltzmann equation for small deflexions (or directly use the Fokker-Planck equation). In that case, we obtain the modified Landau equation (\ref{landaumodif1}) of Appendix \ref{sec_landaumodif}.  This equation does not diverge at small scales since strong collisions
have been accounted for.  We note, however, that a more rigorous kinetic equation should take into account collective effects (and spatial inhomogeneity for attractive interactions) even if their presence is not required to make the integral converge at
large scales.

\subsection{The case $\alpha<\alpha_c$}
\label{sec_inf}

If $\alpha< \alpha_c$ (i.e. $\gamma<\gamma_c$ or $d< 3$ for  a Coulombian or a Newtonian potential), the integral (\ref{pl2}) converges for $k\rightarrow
+\infty$ (small scales) implying that strong collisions are negligible.  On the other hand, it diverges algebraically for $k\rightarrow 0$ (large scales).  In the case of repulsive interactions
(e.g. plasmas), the divergence is regularized by collective effects
which account for Debye shielding. Heuristically, we can introduce a
cut-off at the Debye length and take $k_{min}\sim k_D$. In the case of
attractive interactions (e.g. gravity), the divergence is regularized
by the finite extent of the system. Heuristically, we can  introduce a
cut-off at the Jeans length and take $k_{min}\sim k_J$. Performing the integrals in Eqs. (\ref{pl2}) and (\ref{pl3}) with $k_{min}\sim k_s$, we obtain
\begin{equation}
\label{pl6}
{\cal D}_d=\frac{1}{2(2\pi)^{d-\frac{1}{2}}}\, \frac{1}{3-d-2\alpha}\, \frac{v_m^3 k_s}{\Lambda},\qquad t_R^{bath}\sim  \Lambda t_D.
\end{equation}

We now develop a more precise description (see Appendix \ref{sec_bbgky}).
For repulsive interactions (e.g. plasmas), in the limit $\Lambda\gg 1$ the system is dominated by weak collisions and collective effects (strong collisions and three-body collisions can be neglected). Therefore, its evolution is rigorously described by the Lenard-Balescu equation (II-3). This equation does not present any divergence. In the thermal bath approximation, the diffusion tensor  is given by Eq. (II-39) with
\begin{equation}
\label{pl7}
{\cal D}_d(x)=\frac{1}{2(2\pi)^{d-\frac{1}{2}}}\frac{v_m^3 k_D}{\Lambda}\int_0^{+\infty}  \frac{\kappa^d}{\left\lbrack \kappa^{2-\alpha}+B(x)\right\rbrack^{2}+C(x)^{2}}\, d{\kappa}.
\end{equation}
We note that the large-scale divergence that occurs in the diffusion coefficient (\ref{pl2}) is regularized by Debye shielding. If we make  the Debye-H\"uckel approximation, or consider small velocities $|{\bf v}|\rightarrow 0$, the diffusion tensor
 is given by Eq. (II-45) with\footnote{The value of the integral is
\begin{equation}
\label{pl8b}
\int_0^{+\infty}  \frac{\kappa^d}{(\kappa^{2-\alpha}+1)^{2}}\, d{\kappa}=(1-\alpha-d)\frac{\pi}{(2-\alpha)^2}\frac{1}{\sin\left\lbrack \frac{(1+d)\pi}{2-\alpha}\right \rbrack}.
\end{equation}
}
\begin{equation}
\label{pl8}
{\cal D}^{DH}_d=\frac{1}{2(2\pi)^{d-\frac{1}{2}}}\frac{v_m^3 k_D}{\Lambda}\int_0^{+\infty}  \frac{\kappa^d}{(\kappa^{2-\alpha}+1)^{2}}\, d{\kappa},\qquad  t_{R}^{bath}\sim \frac{\Lambda}{{\int_0^{+\infty}  \frac{\kappa^d\, d{\kappa}}{(\kappa^{2-\alpha}+1)^{2}}}}t_D. \qquad
\end{equation}
If we neglect collective effects and introduce a large-scale cut-off at the Debye length, we get the Landau equation (II-5).  In the thermal bath approach, the diffusion tensor is given by Eq. (II-45) with Eq. (\ref{pl6}). However, the Landau equation with a cut-off at the Debye length is not quantitatively correct since the results depend strongly (algebraically) on the precise value of the large-scale cut-off. It may, however, provide a reasonable approximation of the Lenard-Balescu equation provided that the cut-off is suitably adapted to the situation.

For attractive interactions (e.g. gravity), in the limit $\Lambda\gg 1$ the system is dominated by weak collisions, spatial inhomogeneity, and collective effects (strong collisions and three-body collisions can  be neglected). Therefore, its evolution is rigorously described by the Lenard-Balescu equation  written with angle-action variables \cite{angleaction,kindetail,heyvaerts,newangleaction,aanew}. If we neglect collective effects, it reduces to the Landau equation  written with angle-action variables. These equations do not display any divergence. The large-scale divergence that occurs in the diffusion coefficient (\ref{pl2}) is regularized by the finite extent of the system. If we neglect collective effects, make a local approximation, and introduce a large-scale cut-off at the Jeans length, we get the Vlasov-Landau equation (II-16). In the thermal bath approach, the diffusion tensor is
 given by Eq. (II-45) with Eq. (\ref{pl6}). However, the Vlasov-Landau equation with a cut-off at the Jeans length is not quantitatively correct since the results depend strongly (algebraically) on the precise value of the large-scale cut-off. It may, however, provide a reasonable approximation of the Landau equation  written with angle-action variables provided that the cut-off is suitably adapted to the situation. Finally, we note that collective effects tend to reduce the relaxation time (\ref{pl6}-b) as explained in Appendix \ref{sec_reduction}.

\subsection{The marginal case $\alpha=\alpha_c$}
\label{sec_eq}

If $\alpha=\alpha_c$ (i.e. $\gamma=\gamma_c$ or $d=3$ for  a Coulombian or a Newtonian potential), the integral (\ref{pl2}) diverges logarithmically at small and large scales. This implies that both strong collisions and collective effects (or spatial inhomogeneity for attractive interactions) must be taken into account. However, their influence is weak since the divergence is only logarithmic. We are therefore in a marginal situation.  Heuristically, we can introduce a small-scale cut-off at the Landau length\footnote{We assume $d>1$ because the Landau length
is ill-defined in $d=1$ when $\alpha=\alpha_c$. } and a large-scale cut-off at the Debye length (for repulsive interactions) or at the Jeans length (for attractive interactions). Performing the integrals in Eqs. (\ref{pl2}) and (\ref{pl3}) with $k_{max}\sim k_{L}=\Lambda^{\frac{2}{d-1}}k_s$ and $k_{min}\sim k_s$, we obtain
\begin{equation}
\label{pl10}
{\cal D}_d=\frac{1}{2(2\pi)^{d-\frac{1}{2}}}\, \frac{2}{d-1}\, \frac{v_m^3 k_s\ln\Lambda}{\Lambda},\qquad t_R^{bath}\sim  \frac{\Lambda}{\ln\Lambda} t_D.
\end{equation}

We now develop a more precise description (see Appendix \ref{sec_bbgky}).
For repulsive interactions (e.g. plasmas), in the limit $\Lambda\gg 1$ the evolution of the system is dominated by weak collisions and collective effects. Three-body collisions are negligible. In principle, strong collisions should also be negligible but the divergence of the diffusion coefficient (\ref{pl2}) shows that they are important at small scales. Therefore,
the evolution of the system is rigorously described by Eqs. (\ref{bbgky0}-a) and  (\ref{bbgky0}-b)
with ${\cal T}=0$. These equations do not present any divergence. However, it is difficult to make them
 more explicit. The usual strategy is to consider successively the contribution of collisions with small and large impact parameters, and then connect these two limits. If we ignore collective effects and introduce a
large-scale cut-off at the Debye length $\lambda_{max}=\lambda_D$, we get the Boltzmann equation. This equation does not diverge at small scales since strong collisions are taken into account. This equation is marginally valid since the divergence at large scales is weak (logarithmic). For $\Lambda\gg 1$, weak collisions dominate over strong collisions and we can expand the Boltzmann equation for small deflexions (or directly use the Fokker-Planck equation). In the dominant approximation $\ln\Lambda\gg 1$, we get the Landau equation (II-5) with the diffusion
coefficient (\ref{pl10}) with $\ln\Lambda=\ln(\lambda_{max}/\lambda_L)$ in which the Landau length $\lambda_L$ appears
naturally (see Appendix \ref{sec_lc}). If we ignore strong collisions and introduce a small-scale cut-off
at the Landau length $\lambda_{min}=\lambda_L$, we get the Lenard-Balescu equation. This equation is marginally valid since the divergence at small scales is weak (logarithmic). In the thermal bath approach, the diffusion tensor  is given by Eq. (II-39) with
\begin{equation}
\label{pl11}
{\cal D}_d(x)=\frac{1}{2(2\pi)^{d-\frac{1}{2}}}\frac{v_m^3 k_D}{\Lambda}\int_0^{k_L/k_D}  \frac{\kappa^d}{\left\lbrack \kappa^{(1+d)/2}+B(x)\right\rbrack^{2}+C(x)^{2}}\, d{\kappa}.
\end{equation}
We note that the large-scale divergence that occurs in the Landau diffusion coefficient (\ref{pl2}) is
regularized by Debye shielding. If we make  the Debye-H\"uckel approximation, or consider small
velocities $|{\bf v}|\rightarrow 0$, the diffusion coefficient  is given by Eq. (II-45) with\footnote{The
value of the integral is
\begin{equation}
\label{pl12vf}
\int_0^{k_L/k_D}  \frac{\kappa^d}{\lbrack \kappa^{(1+d)/2}+1\rbrack^{2}}\, d{\kappa}=\frac{2}{d+1}\left\lbrack
\ln\left (1+\Lambda^{\frac{d+1}{d-1}}\right )-\frac{\Lambda^{\frac{d+1}{d-1}}}{1+\Lambda^{\frac{d+1}{d-1}}}\right\rbrack.
\end{equation}
with $k_L/k_D=\Lambda^{2/(d-1)}$.}
\begin{equation}
\label{pl12}
{\cal D}^{DH}_d=\frac{1}{2(2\pi)^{d-\frac{1}{2}}}\frac{v_m^3 k_D}{\Lambda}\int_0^{k_L/k_D}  \frac{\kappa^d}{\lbrack \kappa^{(1+d)/2}+1\rbrack^{2}}\, d{\kappa},\qquad  t_{R}^{bath}\sim \frac{\Lambda}{\int_0^{k_L/k_D}  \frac{\kappa^d\, d{\kappa}}{\lbrack \kappa^{(1+d)/2}+1\rbrack^{2}}}t_D. \qquad
\end{equation}
In the dominant approximation $\ln\Lambda\gg 1$, the Lenard-Balescu diffusion coefficient
${\cal D}_d(x)$ and the Debye-H\"uckel diffusion coefficient ${\cal D}^{DH}_d$ can be approximated
by Eq. (\ref{pl10})  with $\ln\Lambda=\ln(\lambda_D/\lambda_{min})$ in which the Debye length $\lambda_D$ appears 
naturally. In conclusion, in the dominant approximation $\ln\Lambda\gg 1$, the evolution
of the system is rigorously described by the Landau equation with a small scale cut-off at the Landau length and a large-scale cut-off at the Debye length. These cut-offs are not {\it ad hoc} but they  are justified by the above arguments.

For attractive interactions (e.g. gravity), in the limit $\Lambda\gg 1$  the evolution of the system is dominated by weak collisions, collective effects, and spatial inhomogeneity. Three-body collisions are negligible. In principle, strong collisions should also be negligible but the divergence of the diffusion coefficient (\ref{pl2}) shows that they are important at small scales. Therefore, the evolution of
the system is rigorously described by Eqs. (\ref{bbgky1}-a) and  (\ref{bbgky1}-b)  with ${\cal T}=0$. These
equations do not present any divergence.  However, it is difficult to make them more explicit. Again,
the usual strategy is to consider successively the contribution of collisions with small and large impact parameters, and then connect these two limits. If we ignore collective effects and introduce a
large-scale cut-off at the Jeans length $\lambda_{max}=\lambda_J$, we get the Boltzmann equation.  This equation does not diverge at small scales since strong collisions are taken into account. This equation is marginally valid since the divergence at large scales is weak (logarithmic). For $\Lambda\gg 1$, weak collisions dominate over strong collisions and  we can expand the Boltzmann equation for small deflexions (or directly use the Fokker-Planck equation). In the dominant approximation $\ln\Lambda\gg 1$, we get the Vlasov-Landau equation (II-16) with the
diffusion coefficient (\ref{pl10}) with $\ln\Lambda=\ln(\lambda_{max}/\lambda_L)$ in which the Landau
length $\lambda_L$ appears naturally (see Appendix \ref{sec_lc}). If we ignore strong collisions and introduce a
small-scale cut-off at the Landau length $\lambda_{min}=\lambda_L$, we get the Lenard-Balescu equation written with angle-action variables \cite{angleaction,kindetail,heyvaerts,newangleaction,aanew}. This equation is marginally valid since the divergence at small scales is weak (logarithmic). If we neglect collective effects, it reduces to the Landau equation written with angle-action variables. These equations do not  diverge at large scales since they take into account the finite extent of the system.   If we neglect collective effects and make a local approximation (which is justified by the fact that the divergence of the diffusion coefficient at large scales is weak), we get the Vlasov-Landau equation (II-16). This equation presents a logarithmic divergence at large scales ($k\rightarrow 0$). Since this divergence is cured by the finite extent of the system, it is natural to introduce a large-scale cut-off at the Jeans scale. Since this procedure is essentially heuristic, the precise value of the large-scale cut-off is not known. However, in the dominant approximation $\ln\Lambda\gg 1$, the precise value of the numerical factor
is not important since it appears in a logarithmic factor and we finally obtain the  Vlasov-Landau
equation (II-16) with the diffusion coefficient (\ref{pl10})
with $\ln\Lambda=\ln(\lambda_{J}/\lambda_{min})$. In conclusion, in the dominant approximation $\ln \Lambda\gg 1$, the evolution of the system is reasonably well described by the Vlasov-Landau equation (II-16) with a small scale cut-off at the Landau length and a large-scale cut-off at the Jeans length. This equation is not exact (the rigorous equation is the Lenard-Balescu equation written in angle-action variables with a small-scale cut-off at the Landau length justified by the previous arguments) but it may be used as a simplified kinetic equation. Finally, we note that collective effects tend to reduce the relaxation time (\ref{pl10}-b) as explained in Appendix \ref{sec_reduction}.

\subsection{Existence of quasi stationary states}
\label{sec_qss}

The previous scalings of the relaxation time agree with those obtained by Gabrielli {\it et al.} \cite{gjm}
from the Chandrasekhar binary collision theory\footnote{Their approach assumes that
collective effects and spatial inhomogeneity can be neglected.
Our approach is more general since it can take these effects
into account as explained previously.}. For $\alpha<\alpha_c$, the Lenard-Balescu equation is rigorously valid and the relaxation
time scales as $\Lambda t_D$. For $\alpha=\alpha_c$, the Lenard-Balescu equation is marginally valid and the relaxation time scales as $(\Lambda/\ln\Lambda)t_D$. This corresponds to the usual situation encountered in plasma physics and stellar dynamics ($\alpha=0$) since the dimension of space $d=3$ is critical for Coulombian and Newtonian interactions. For $\alpha>\alpha_c$, the Lenard-Balescu equation is not valid and the  relaxation time scales as $\Lambda^{\frac{1-\alpha}{d-2+\alpha}}t_D$.

For $\alpha<1$ (corresponding to $\gamma<d-1$), the relaxation time diverges when $\Lambda\rightarrow +\infty$ (for $\alpha<\alpha_c$, the relaxation time increases linearly with $\Lambda$ and, for $\alpha_c\le\alpha<1$, it increases less rapidly than $\Lambda$). In that case, the Vlasov regime becomes infinite in the thermodynamic limit and quasi stationary states (QSS) may form
as a result of violent relaxation \cite{lb}. On the other hand, for $\alpha>1$ (corresponding to $\gamma>d-1$), the relaxation time tends to zero when $\Lambda\rightarrow +\infty$. In that case, as 
emphasized by  Gabrielli {\it et al.} \cite{gjm}, no QSS can form (except if the potential includes a sufficiently
large soft core). In the intermediate case $\alpha=1$, the relaxation time is independent on $\Lambda$.

{\it Remark 3:} If we define the mean free path by $\lambda\sim v_m t_R$, we find that $\lambda\sim \Lambda^{(1-\alpha)/(d-2+\alpha)}\lambda_s$ for $\alpha_c<\alpha<2$, $\lambda\sim (\Lambda/\ln\Lambda)\lambda_s$ for $\alpha=\alpha_c$, and $\lambda\sim \Lambda \lambda_s$ for $\alpha<\alpha_c$. For $\alpha<1$, we find that $\lambda\gg\lambda_s$ when $\Lambda\gg 1$ so that the mean free path is much larger than the Debye length or the Jeans length. This is another manner to understand why the Vlasov equation becomes
exact when $\Lambda\rightarrow +\infty$.

{\it Remark 4:} For specific power-law potentials, the dynamics may be particular. For $\gamma=-2$ (harmonic 
potential)  there is no (violent or slow) relaxation and the  system oscillates permanently as discovered 
by Newton \cite{newton,chandranewton,lblb}.

\section{Coulombian plasmas}
\label{sec_plasmas}

For Coulombian plasmas, the Fourier transform of the potential of interaction is
\begin{equation}
\label{tp1}
(2\pi)^d\hat{u}(k)=\frac{S_d e^2}{m^2k^2},\qquad \eta(k)=-\frac{k_D^2}{k^{2}},
\end{equation}
and the Fourier transform of the Debye-H\"uckel potential is
\begin{equation}
\label{tp1b}
(2\pi)^d\hat{u}_{DH}(k)=\frac{S_d e^2}{m^2}\frac{1}{k^2+k_D^2},\qquad \eta_{DH}(k)=-\frac{k_D^2}{k^{2}+k_D^2}.
\end{equation}
In physical space, we have $u(r)=(e^2/m^2) r^{-1}$ and $u_{DH}(r)=(e^2/m^2) e^{-k_D r}/r$
in $d=3$,  $u(r)=-(e^2/m^2) \ln(r)$ and $u_{DH}(r)=(e^2/m^2) K_0(k_D r)$
in $d=2$, $u(x)=-(e^2/m^2)|x|$ and $u_{DH}(x)=(e^2/m^2 k_D) e^{-k_D |x|}$ in $d=1$.

\subsection{3D Coulombian plasmas}
\label{sec_tp}

If we neglect strong collisions and collective effects, the evolution of the system is described by the Landau equation (II-5). In the thermal bath approach, the diffusion tensor
is given by Eq. (II-45) with
\begin{equation}
\label{tp1gr}
{\cal D}_3=\frac{1}{2(2\pi)^{\frac{5}{2}}}\frac{v_m^3}{n}k_D^4\int_0^{+\infty}  \frac{dk}{k}.
\end{equation}
The integral diverges logarithmically at small and large scales. This implies that both strong collisions and collective effects (Debye shielding) must be taken into account. However, their effect is weak since the divergence is logarithmic (we are in the marginal case of Sec. \ref{sec_eq}). If we introduce a small-scale cut-off $k_{max}=k_{L}=\Lambda k_D$ at  the Landau length at which binary collisions become strong and a large-scale cut-off $k_{min}=k_D$ at the Debye length at which the interaction is shielded, we get
 \begin{equation}
\label{tp2}
{\cal D}_3=\frac{1}{2(2\pi)^{\frac{5}{2}}}v_m^3 k_D\frac{\ln\Lambda}{\Lambda},
\end{equation}
where $\Lambda= n k_D^{-3}\gg 1$ represents the number of electrons in the Debye sphere. In plasma physics, $\ln\Lambda$ is called the Coulomb logarithm.

Strong collisions are taken into account in the Boltzmann equation. This corresponds to the two-body encounters (or impact) theory. In this equation, the integral over the impact parameter does not diverge at small scales. Therefore, the correct treatment of strong collisions regularizes the logarithmic divergence that appears at small scales in the Landau equation. However, for a Coulombian potential in $d=3$, the integral over the impact parameter diverges logarithmically at large scales.  Therefore, the Boltzmann equation  is marginally valid provided that a large-scale cut-off is introduced
at the Debye length $\lambda_{max}=\lambda_D$ (see the next argument). If we expand the Boltzmann equation for small deflexions (or directly use the Fokker-Planck equation), and consider the dominant approximation $\ln\Lambda \gg 1$, we obtain the Landau equation (II-5) with the diffusion
coefficient (\ref{tp2}) with $\ln\Lambda=\ln(\lambda_{max}/\lambda_L)$ in which the Landau length
$\lambda_L$ appears naturally (see Appendix \ref{sec_lc}).

Collective effects are taken into account in the Lenard-Balescu equation (II-3). This corresponds to the wave  theory. In this equation, the integral over the wavenumber does not diverge at large scales. Therefore, the correct treatment of collective effects regularizes the logarithmic divergence that appears at large scales in the Landau equation. However, for a Coulombian potential in $d=3$, the integral over the wavenumber diverges logarithmically at small scales. Therefore, the  Lenard-Balescu equation is marginally valid provided that a small-scale cut-off is introduced
 at the Landau length $\lambda_{min}=\lambda_L$  (see the previous argument).  In the thermal bath approach, the diffusion tensor is given by Eq. (II-39) with\footnote{The value of the integral is \begin{equation}
\label{tp4bfoot}
\int_0^{k_L}  \frac{k^3}{\left\lbrack k^2+B(x)k_D^2\right\rbrack^{2}+C(x)^{2}k_D^4}\, d{k}=\ln\Lambda+\frac{1}{4}\ln\left\lbrack \frac{\left (1+\frac{B(x)}{\Lambda^2}\right )^2+\frac{C(x)^2}{\Lambda^4}}{B(x)^2+C(x)^2}\right\rbrack-\frac{B(x)}{2|C(x)|}\left\lbrack\frac{\pi}{2}
-\arctan\frac{B(x)}{|C(x)|}\right\rbrack.
\end{equation}
}
\begin{equation}
\label{tp3}
{\cal D}_3(x)=\frac{1}{2(2\pi)^{\frac{5}{2}}}\frac{v_m^3}{n}k_D^4\int_0^{k_L}  \frac{k^3}{\left\lbrack k^2+B(x)k_D^2\right\rbrack^{2}+C(x)^{2}k_D^4}\, d{k}.
\end{equation}
This integral converges at large scales ($k\rightarrow 0$) due to Debye shielding. For small velocities $|{\bf v}|\rightarrow 0$, the diffusion coefficient is given by (II-45) with\footnote{The value of the integral is \begin{equation}
\label{tp4b}
\int_0^{k_L}  \frac{k^3}{\left ( k^2+k_D^2\right )^{2}}\, d{k}=\frac{1}{2}\left\lbrack \ln(1+\Lambda^2)-\frac{\Lambda^2}{1+\Lambda^2}\right \rbrack.
\end{equation}}
\begin{equation}
\label{tp4}
{\cal D}_3^{DH}={\cal D}_3(0)= \frac{1}{2(2\pi)^{\frac{5}{2}}}\frac{v_m^3}{n}k_D^4\int_0^{k_L}  \frac{k^3}{\left ( k^2+k_D^2\right )^{2}}\, d{k}.
\end{equation}
This corresponds to the Debye-H\"uckel approximation.  In  the dominant approximation $\ln\Lambda\gg 1$,
the Lenard-Balescu diffusion coefficient (\ref{tp3}) and  Debye-H\"uckel diffusion coefficient (\ref{tp4})
reduce to the Landau diffusion coefficient (\ref{tp2}) with $\ln\Lambda=\ln(\lambda_D/\lambda_{min})$ in which the Debye length appears naturally.  This shows that the Landau equation with a large-scale cut-off at the Debye length provides a very good approximation of the Lenard-Balescu equation in $d=3$. This implies that dynamical screening is 
negligible\footnote{This is true for the Maxwell distribution. This is not true anymore 
for distributions for which the large velocity population plays a crucial role as compared to the bulk of 
the distribution.} in $d=3$. Indeed, in the dominant approximation, we can neglect the velocity effects encapsulated in the dielectric function $\epsilon({\bf k},{\bf k}\cdot {\bf v})$ and use the Debye-H\"uckel potential accounting for static screening.

For a Coulombian plasma in $d=3$ (marginal case), the impact theory and the wave theory are two approximate theories that are complementary to each other. Collective interactions between charges
are not included in the impact theory implying that the integrand in the Boltzmann equation is valid only for impact parameters sufficiently smaller than the Debye length ($\lambda\ll\lambda_D$). On the other hand, the curvature of orbits at small impact parameters is not included in the wave theory implying that the integrand in the Lenard-Balescu equation is valid only for wavelengths sufficiently  larger than the Landau length ($\lambda\gg\lambda_L$).
For $\Lambda\gg 1$, the range of validity of these two theories greatly overlaps in the region $\lambda_L\ll\lambda\ll\lambda_D$ corresponding to the domain of validity of the Landau equation.
A possibility to unify these approaches and determine precisely the numerical factor in the Coulomb logarithm was first recognized by Hubbard \cite{hubbard2} and subsequently  developed  by Kihara and Aono \cite{ka} and Gould and DeWitt \cite{gould} among others. Roughly speaking, the idea is to sum the Boltzmann ($0<\lambda<\lambda_D$) and Lenard-Balescu  ($\lambda_L<\lambda<+\infty$) equations and subtract the Landau equation ($\lambda_L<\lambda<\lambda_D$). This leads to a fully convergent kinetic equation. It has the form of the Landau equation in which the Coulomb logarithm is exactly given by $\ln\Lambda-2\gamma+\ln 2$ where $\gamma=0.5772$ is Euler's constant. If we neglect the constant term as compared to $\ln\Lambda$ (dominant approximation), this procedure {\it justifies} using the Landau equation with a small-scale cut-off at the Landau length and a large-scale cut-off
at the Debye length.

{\it Conclusion:} The evolution of the system as a whole is described by the Lenard-Balescu equation (II-3) with a cut-off at the Landau length (this cut-off is not {\it ad hoc} but is justified by the correct treatment of strong collisions). The relaxation of a test particle in a thermal bath is described by the Fokker-Planck equation  (II-36) with the diffusion tensor given by Eqs. (II-39) and (\ref{tp3}). In the dominant approximation $\ln\Lambda\gg 1$, the Lenard-Balescu equation can be replaced  by the Landau equation (II-5) with a cut-off at the Debye length (this cut-off is not {\it ad hoc} but is justified by the correct treatment of collective effects). In that case, the evolution of the system as a whole is described  by the Landau equation (II-5) or (II-9) with $K_3=(2\pi e^4/m^3)\ln(\lambda_{D}/\lambda_L)$. The relaxation of a test particle in a thermal bath in described by the Fokker-Planck equation  (II-36) with the diffusion tensor
given by Eqs. (II-47) and (II-48) where ${\cal D}_3$ given by Eq. (\ref{tp2}). Using the results of Sec. \ref{sec_powerlaw}, we find that the relaxation time is given by
\begin{eqnarray}
t_R\sim \frac{m^2v_m^3}{ne^4 \ln\left (\frac{m^{3/2}v_m^3}{e^3n^{1/2}}\right )}\sim \frac{\Lambda}{\ln \Lambda} t_D,
\label{tp5}
\end{eqnarray}
where $t_D\sim \omega_P^{-1}$ is the dynamical time.  A short historic of the early development
of the kinetic theory of 3D Coulombian plasmas with many references
is given in Paper II.

\subsection{2D Coulombian plasmas}
\label{sec_bp}

If we neglect strong collisions and collective effects, the evolution of the system is described
by the Landau equation (II-5). In the thermal bath approach, the diffusion tensor is given by Eq. (II-45) with
\begin{equation}
\label{bp1}
{\cal D}_2=\frac{1}{2(2\pi)^{\frac{3}{2}}}\frac{v_m^3}{n}k_D^4\int_{0}^{+\infty}  \frac{dk}{k^2}.
\end{equation}
This integral converges at small scales implying that strong collisions are negligible. However, it diverges rapidly (linearly) at large scales implying that collective effects (Debye shielding) are important. If we introduce a large-scale cut-off $k_{min}=k_D$ at the Debye length, we obtain
\begin{equation}
\label{bp2}
{\cal D}_2=\frac{1}{2(2\pi)^{\frac{3}{2}}}v_m^3 k_D\frac{1}{\Lambda},
\end{equation}
where $\Lambda= n k_D^{-2}\gg 1$ represents the number of electrons in the Debye disk.

Collective effects are taken into account in the Lenard-Balescu equation (II-3). This equation is rigorously valid for a Coulombian potential in $d=2$ (see Sec. \ref{sec_inf}). The large-scale divergence that occurs in the Landau equation is regularized by Debye shielding. In the thermal bath approach, the diffusion tensor is
given by Eq. (II-39) with
\begin{equation}
\label{bp3}
{\cal D}_2(x)=\frac{1}{2(2\pi)^{\frac{3}{2}}}\frac{v_m^3}{n}k_D^4\int_0^{+\infty}  \frac{k^2}{\left\lbrack {k^2}+B(x)k_D^2\right\rbrack^{2}+C(x)^{2}k_D^4}\, d{k}.
\end{equation}
This function is perfectly well-defined. For small velocities $|{\bf v}|\rightarrow 0$, the diffusion coefficient is given
 by Eq. (II-45) with
\begin{equation}
\label{bp4}
{\cal D}_2^{DH}={\cal D}_2(0)=\frac{1}{2(2\pi)^{\frac{3}{2}}}\frac{v_m^3}{n}k_D^4\int_0^{+\infty}  \frac{k^2}{(k^2+k_D^2)^2}\, d{k}=\frac{1}{2(2\pi)^{\frac{3}{2}}}v_m^3 k_D\frac{\pi}{4\Lambda}=\frac{\pi}{4}{\cal D}_2.
\end{equation}
This corresponds to the Debye-H\"uckel approximation. For $|{\bf v}|\rightarrow +\infty$, we
can use the asymptotic results given in Sec. IV.D of Paper I. The diffusion coefficient $D_{\perp}(v)$ is given
by Eq. (II-53-d) with ${\cal D}_2$ replaced by ${\cal D}_2(0)$ and the diffusion coefficient $D_{\|}(v)$ is given
by Eq. (II-53-c) with ${\cal D}_2$ replaced by ${\cal D}_2(1)$ where
\begin{equation}
\label{bp5}
{\cal D}_2(1)=\frac{1}{2(2\pi)^{\frac{3}{2}}}\frac{v_m^3}{n}k_D^4\int_0^{+\infty}  \frac{k^2}{\left\lbrack k^2+B(1)k_D^2\right\rbrack^{2}+C(1)^{2}k_D^4}\, d{k}=\frac{1}{2(2\pi)^{\frac{3}{2}}}v_m^3 k_D\frac{1.45804}{\Lambda}=1.45804{\cal D}_2.
\end{equation}
We note that the Debye-H\"uckel approximation is not correct at large velocities (compare Eqs. (\ref{bp4})
and (\ref{bp5})). Therefore, dynamical screening is important for 2D plasmas, contrary to 3D plasmas in the dominant approximation. This makes the dimension $d=2$ particularly interesting. On the other hand, the Lenard-Balescu diffusion coefficient (\ref{bp3}) and the Debye-H\"uckel diffusion coefficient (\ref{bp4}) are different from the Landau diffusion coefficient (\ref{bp2}). Therefore, the Lenard-Balescu
equation cannot be rigorously approximated by the Landau equation with a large-scale cut-off at the Debye length. However, comparing Eqs. (\ref{bp2}), (\ref{bp4}) and (\ref{bp5}), we see that the discrepancy  is not dramatic since the velocity dependence of the diffusion coefficient is the same (at least asymptotically) and the value of the prefactor is only slightly different ($\pi/4=0.785398$ and $1.45804$ instead of $1$). Therefore, the Landau equation may be used as a rough approximation of the Lenard-Balescu equation by ``adapting'' the value of the large scale cut-off.

{\it Conclusion:} The evolution of the system as a whole is described by the Lenard-Balescu equation (II-3). The relaxation of a test particle in a thermal bath in described by the Fokker-Planck equation (II-36) with the diffusion tensor given by Eqs. (II-39) and (\ref{bp3}).  The Landau equation (II-5) with a cut-off at the Debye length is not rigorously equivalent to the Lenard-Balescu equation in $d=2$ but the discrepancy is not dramatic. If we use this approximate theory, the evolution of the system as a whole is given by the
Landau equation (II-5) or (II-9) with $K_2=2\pi e^4/m^3 k_D$, or $K_2^{DH}=\pi^2e^4/2m^3k_D$ in the
Debye-H\"uckel approximation. The evolution of a test particle in a thermal bath is described by the Fokker-Planck equation (II-36) with the diffusion tensor given by Eqs. (II-51) and (II-52) where ${\cal D}_2$ given by Eq. (\ref{bp2}), or by Eq. (\ref{bp4}) in the Debye-H\"uckel approximation. Using the results of Sec. \ref{sec_powerlaw}, we find that the relaxation time is given by
\begin{eqnarray}
t_R\sim \frac{m^{3/2}v_m^2}{n^{1/2}e^3}\sim \Lambda t_D,
\label{bp6}
\end{eqnarray}
where $t_D\sim \omega_P^{-1}$ is the dynamical time.

A kinetic theory of 2D Coulombian plasmas has been developed  by Benedetti {\it et al.} \cite{benedetti} for the cut-off Coulombian potential and by Chavanis \cite{landaud} for the true Coulombian potential. However, collective effects are treated in an approximate manner in Appendix C of that paper. The present treatment is more satisfactory.

\subsection{1D Coulombian plasmas}
\label{sec_daw}

If we neglect strong collisions and collective effects, the evolution of the system is described by the Landau equation (II-5). In the thermal bath approach, the diffusion coefficient is given
by Eqs. (II-46) and (II-54).  This leads to
\begin{equation}
\label{daw1}
D({v})=\frac{1}{(2\pi)^{{1}/{2}}}\frac{v_m^3}{n}k_D^4  e^{-\frac{1}{2}\beta m v^{2}} \int_{0}^{+\infty}  \frac{dk}{k^3}.
\end{equation}
This integral converges at small scales implying that  strong collisions are negligible. However, it diverges rapidly (quadratically) at large scales implying that collective effects (Debye shielding) are important. If we introduce a large-scale cut-off $k_{min}=k_D$ at the Debye length, we obtain
\begin{equation}
\label{daw2}
D({v})=\frac{k_D}{2(2\pi)^{1/2}(\beta m)^{3/2}\Lambda}e^{-\frac{1}{2}\beta m v^{2}},
\end{equation}
where $\Lambda= n k_D^{-1}\gg 1$ represents the number of electrons in the Debye segment.

Collective effects are described by the Lenard-Balescu equation (II-3). This equation is rigorously valid for a Coulombian potential in $d=1$ (see Sec. \ref{sec_inf}). The large-scale divergence that occurs in the Landau equation is regularized by Debye shielding. In the thermal bath approach, the diffusion  coefficient is
given by Eqs. (II-43) and (II-44) leading to
\begin{equation}
\label{daw3}
D(v)=\frac{1}{(2\pi)^{{1}/{2}}}\frac{v_m^3}{n}k_D^4e^{-\frac{1}{2}\beta m v^{2}}\int_0^{+\infty}  \frac{k}{\left\lbrack {k^{2}}+B\left (\sqrt{\frac{\beta m}{2}}\, v\right )k_D^2\right\rbrack^{2}+C\left (\sqrt{\frac{\beta m}{2}}\, v\right )^{2}k_D^4}\, d{k}.
\end{equation}
The integral is perfectly well-defined and can be calculated analytically. Using the identity
\begin{equation}
\label{daw4}
\Phi_1(y)\equiv \int_0^{+\infty}\frac{\kappa}{(\kappa^2+y)^2+1}\, d\kappa=\frac{\pi}{4}-\frac{1}{2}\arctan(y)=\frac{1}{2}{\rm arcot}(y),
\end{equation}
we obtain
\begin{equation}
\label{daw5}
D(v)=\frac{k_D}{2\pi (\beta m)^2 \Lambda |v|}{\rm arcot} \left (\frac{1-\sqrt{2\beta m} v e^{-\frac{1}{2}\beta m v^2}\int_0^{\sqrt{\frac{\beta m}{2}}v} e^{y^2}\, dy}{\sqrt{\frac{\beta \pi m}{2}}|v| e^{-\frac{1}{2}\beta m v^2}}\right ).
\end{equation}
Since ${\rm arcot}(x)\sim 1/x$ for $x\rightarrow +\infty$ and  ${\rm arcot}(x)\rightarrow \pi$ for $x\rightarrow -\infty$, we have the asymptotic behaviors
\begin{equation}
\label{daw6}
D(0)=\frac{k_D}{2(2\pi)^{1/2}(\beta m)^{3/2}\Lambda},\qquad D(v)\sim_{\pm\infty} \frac{k_D}{2(\beta m)^{2}\Lambda |v|}.
\end{equation}
We note that, for $v=0$, the exact diffusion coefficient given by Eq. (\ref{daw6}-a),  coincides with the expression (\ref{daw2}) obtained from the Landau equation by introducing a large-scale cut-off at the Debye length. However, this is essentially coincidental. Indeed, the exact diffusion coefficient for $v=0$ may be written as
\begin{equation}
\label{daw7}
D(0)=\frac{1}{(2\pi)^{{1}/{2}}}\frac{v_m^3}{n}k_D^4\int_0^{+\infty}  \frac{k}{(k^2+k_D^2)^{2}}\, d{k}.
\end{equation}
This corresponds to the Debye-H\"uckel approximation. It turns out that the value of this integral is the same as the value of the integral (\ref{daw1}) with a large-scale cut-off at $k_D$. However, this is not generally true. Collective effects are not equivalent to introducing a large-scale cut-off at $k_D$ in the Landau diffusion coefficient, even for $v=0$ (see, e.g., Sec. \ref{sec_bp}). On the other hand, the diffusion coefficients (\ref{daw2}) and (\ref{daw5}) are totally different at large velocities $|v|\rightarrow +\infty$ [compare Eqs. (\ref{daw2}) and (\ref{daw6}-b)]. Indeed, the exact diffusion coefficient (\ref{daw5}) decays algebraically as $|v|^{-1}$ while the diffusion coefficient  (\ref{daw2}) based on the Landau equation with a large-scale cut-off at the Debye length decays exponentially rapidly as $e^{-\beta m v^2/2}$. This is a crucial difference. This clearly shows that, in $d=1$, the Landau equation with a cut-off at the Debye length gives wrong results (except for low velocities). Collective effects must therefore be properly taken into account by using the Lenard-Balescu equation.

{\it Conclusion:} The evolution of the system as a whole is described by the Lenard-Balescu equation (II-3).
Actually, this equation reduces to $\partial f/\partial t=0$ in $d=1$ (see Eq. (II.11)). Therefore, the collision term vanishes
at the order $1/\Lambda$. If it does not vanish at
the order $1/\Lambda^2$, the relaxation
time of the system as a whole is given by
\begin{eqnarray}
t_R^{whole}\sim \frac{n^{1/2}m^{3/2} v_m^2}{e^3} \sim \Lambda^2 t_D> \Lambda t_D.
\label{daw9}
\end{eqnarray}
The relaxation of a test particle in a thermal bath is given by the Fokker-Planck equation (II-42) with the diffusion coefficient (\ref{daw5}). Using the results of Sec. \ref{sec_powerlaw}, we find that the relaxation time of a test particle in a thermal bath is given by
\begin{eqnarray}
t_R^{bath}\sim \frac{m v_m}{e^2}\sim \Lambda t_D,
\label{daw8}
\end{eqnarray}
where $t_D\sim \omega_P^{-1}$ is the dynamical time. We note that the relaxation time is independent on the density.

As discussed in Paper II, in $d=1$ the relaxation time of a test particle in a thermal bath is different from the relaxation time of the system as a whole. Since the distribution of the field particles does not change on a timescale of the order $\Lambda t_D$, we can consider the relaxation of a test particle in an out-of-equilibrium bath (see Sec. IV.C of Paper II). This relaxation process is described by
the Fokker-Planck equation (II-63) with a diffusion coefficient given by Eq. (II-58) and (II-61). Using the identity (\ref{daw4}), it can be written as
\begin{equation}
\label{daw10} D(v)=\frac{e^2f(v)}{\pi m|f'(v)|}{\rm arcot} \left (-\frac{1}{\pi |f'(v)|}{\cal P}\int_{-\infty}^{+\infty}\frac{f'(u)}{u-v}\, dv\right ).
\end{equation}
For the isothermal distribution, we recover Eq. (\ref{daw5}). For the waterbag distribution, using Eq. (II-64), we get
\begin{equation}
\label{daw11} D(v)= \frac{e^2v_m}{2m}\left\lbrack 1-\left (\frac{v}{v_m}\right )^2\right\rbrack,
\end{equation}
for $-v_m\le v\le v_m$ and $D(v)=0$ otherwise. When collective effects are neglected, the diffusion coefficient is given
by Eq. (II-65). However, it involves a divergent integral as in Eq. (\ref{daw1}) for the isothermal distribution, so this approximation is not reliable.

A one component 1D plasma model with a neutralizing background\footnote{The equilibrium properties of a two-components  1D plasma where the electron and 
ions move freely has been studied by
Eldridge and Feix \cite{ef1}. This corresponds to the Lenard-Prager model \cite{lenardprager,pragerlenard}.} has been studied numerically  by Dawson \cite{dawsonfirst,dawson}. This is one of the first models in physics where numerical simulations have been made to test non-equilibrium properties of particles in interaction. The kinetic theory of this model, taking collective effects into account, has been developed 
by Eldridge and Feix \cite{ef2,feix}\footnote{We independently obtained the same results in Sec. 2.8.3 of \cite{hb2}.}. For what concerns the evolution of the system as a whole, they found that the Lenard-Balescu collision term vanishes in 1D so that the relaxation time $t_R$ is strictly larger that $\Lambda t_D$ (numerical simulations
show that it scales like $\Lambda^2 t_D$ which corresponds to the next order in the expansion of the
BBGKY hierarchy). On the other hand, considering the relaxation of a test particle in a thermal bath, they found that the evolution of the distribution function is governed by the Fokker-Planck
equation (II-42) with the diffusion coefficient (\ref{daw5}). In that case, the relaxation time scales 
like $t_R^{bath}\sim \Lambda t_D$. Recently, the Dawson model has been re-investigated
by Sano and Kitahara \cite{sk}. The present results complete their theoretical analysis.

\section{Self-gravitating systems}
\label{sec_sg}

For self-gravitating systems, the Fourier transform of the potential of interaction is
\begin{equation}
\label{sg1}
(2\pi)^d\hat{u}(k)=-\frac{S_d G}{k^2},\qquad \eta(k)=\frac{k_J^2}{k^{2}},
\end{equation}
and the Fourier transform of the  Debye-H\"uckel potential is
\begin{equation}
\label{sg2}
(2\pi)^d\hat{u}_{DH}(k)=-\frac{S_d G}{k^2-k_J^2},\qquad \eta_{DH}(k)=\frac{k_J^2}{k^{2}-k_J^2}.
\end{equation}
In physical space, we have $u(r)=-G/r$ and $u_{DH}(r)=-G \cos(k_J r)/r$ in $d=3$, $u(r)=G\ln r$
 and $u_{DH}(r)=\frac{\pi}{2}G Y_0(k_J r)$
in $d=2$, $u(x)=G|x|$ and $u_{DH}(x)=(G/k_J) \sin(k_J |x|)$ in $d=1$.

\subsection{3D self-gravitating systems}
\label{sec_sgthree}

If we neglect strong collisions, collective effects, and make a local approximation, the evolution of the system is described by the Vlasov-Landau equation (II-16). In the thermal bath approach, the diffusion tensor is given
by Eq. (II-45) with
\begin{equation}
\label{sgthree1}
{\cal D}_3=\frac{1}{2(2\pi)^{\frac{5}{2}}}\frac{v_m^3}{n}k_J^4\int_0^{+\infty}  \frac{dk}{k}.
\end{equation}
This integral diverges logarithmically at small and large scales. This means that both strong collisions and spatial inhomogeneity must be taken into account. However, their effect is weak since the divergence is only logarithmic (we are in the marginal case of Sec. \ref{sec_eq}).  If we
introduce a small-scale cut-off $k_{max}=k_{L}=\Lambda k_J$ at the Landau length at which binary collisions become strong and a large-scale cut-off $k_{min}=k_J$ at the Jeans length which represents the typical size of the system, we get
\begin{equation}
\label{sgthree2}
{\cal D}_3=\frac{1}{2(2\pi)^{\frac{5}{2}}}v_m^3 k_J\frac{\ln \Lambda}{\Lambda},
\end{equation}
where $\Lambda=n k_J^{-3}\sim N\gg 1$ represents the total number of stars in the cluster.

Strong collisions are taken into account in the Boltzmann equation. This corresponds to the two-body encounters (or impact) theory. In this equation, the integral over the impact parameter does not diverge at small scales. Therefore, the correct treatment of strong collisions regularizes the logarithmic divergence that appears at small scales in the Landau equation. However, for a Newtonian potential in $d=3$, the integral over the impact parameter diverges logarithmically at large scales.  Therefore, the Boltzmann equation is marginally valid for a Newtonian potential in $d=3$ provided that a large-scale cut-off is introduced
at the Jeans length  $\lambda_{max}=\lambda_J$
(see the next argument). Since weak collisions dominate over strong collisions,  we can expand the Boltzmann
equation for small deflexions (or directly use the Fokker-Planck equation). In the dominant approximation
$\ln \Lambda\gg 1$, we obtain the Landau equation (II-16) with the diffusion coefficient (\ref{sgthree2})
with $\ln\Lambda=\ln(\lambda_{max}/\lambda_L)$ in which the Landau length $\lambda_L$ appears naturally
 (see Appendix \ref{sec_lc}).

Spatial inhomogeneity and collective effects are taken into account in the Lenard-Balescu equation  written with angle-action variables \cite{angleaction,kindetail,heyvaerts,newangleaction,aanew}. This corresponds to the wave  theory.  If we neglect collective effects, we get the Landau equation written with angle-action variables. In these equations,  the integral over the wavenumber does not diverge at large scales since  the finite extent of the system is taken into account. Therefore, the correct treatment of spatial inhomogeneity regularizes the logarithmic divergence that appears at large scales in the Landau equation. However, for a Newtonian potential in $d=3$, the integral over the wavenumber diverges logarithmically at small scales. Therefore the Lenard-Balescu and Landau equations written with angle-action variables are marginally valid for a Newtonian
 interaction in $d=3$ provided that a small-scale cut-off is introduced at the
Landau length $\lambda_{min}=\lambda_L$
(see the previous argument). If we neglect collective effects, make a local approximation, and introduce
a large scale cut-off at the Jeans length, we get the Vlasov-Landau
equation (II-16) with the diffusion coefficient (\ref{sgthree2}) with $\ln\Lambda=\ln(\lambda_{J}/\lambda_{min})$. In the dominant approximation $\ln \Lambda\gg 1$, this equation does not depend on the precise value of the large-scale cut-off since it appears in a logarithmic factor. This suggests that the local approximation is reasonable and that the Vlasov-Landau equation (II-3) provides a good approximation of the true dynamics of a stellar system in $d=3$.

For a stellar system in $d=3$ (marginal case), the impact theory and the wave theory are two approximate theories that are complementary to each other. Collective interactions between stars
are not included in the impact theory implying that the integrand in the Boltzmann equation is valid only for impact parameters sufficiently smaller than the Jeans length ($\lambda\ll\lambda_J$). On the other hand, the curvature of orbits at small impact parameters is not included in the wave theory implying that the integrand in the Lenard-Balescu equation with angle-action variables is valid only for wavelengths sufficiently  larger than the Landau length ($\lambda\gg\lambda_L$). For $\Lambda\gg 1$, the range of validity of these two theories greatly overlaps in the region $\lambda_L\ll\lambda\ll\lambda_J$ corresponding to the domain of validity of the Vlasov-Landau equation. Combining these approaches, we can motivate (but not rigorously justify) using  the Vlasov-Landau equation with a small-scale cut-off at the Landau length and a large-scale cut-off at the Jeans scale. A better kinetic equation is the Lenard-Balescu equation written with angle-action variables with a small-scale cut-off at the Landau length.

{\it Conclusion:} The evolution of the system as a whole is described  by Lenard-Balescu 
equation (taking collective effects into account) or by
the Landau equation (neglecting collective effects) written with angle-action variables  
with a small-scale cut-off at the Landau length \cite{angleaction,kindetail,heyvaerts,newangleaction,aanew}. The evolution of a test particle in a thermal bath is described by the Fokker-Planck equation written with angle-action variables with a small-scale cut-off at the Landau length.  If we neglect collective effects,  make a local approximation, and introduce  a large scale cut-off at the Jeans length, the evolution of the system as a whole is described by the Vlasov-Landau equation (II-16) or (II-17)
with $K_3=2\pi m G^2 \ln \Lambda$.  The relaxation of a test particle in a thermal bath is described by
the Fokker-Planck equation (II-56) with the diffusion tensor given by Eqs. (II-47) and (II-48) where ${\cal D}_3$ is given by Eq. (\ref{sgthree2})\footnote{For self-gravitating systems, a pure isothermal distribution $f({\bf r},{\bf v})=Ae^{-\beta m (v^2/2+\Phi({\bf r}))}$ leads to a density profile with an infinite mass. Therefore, a pure isothermal distribution function cannot hold in the whole cluster. We can, however, apply the Fokker-Planck theory in the core of the system
which is approximately isothermal.}. Using the results of Sec. \ref{sec_powerlaw}, we find that the relaxation time is given by
\begin{eqnarray}
t_R\sim \frac{v_m^3}{ n m^2 G^2\ln\left (\frac{v_m^3}{n^{1/2}m^{3/2}G^{3/2}}\right )}\sim \frac{N}{\ln N} t_D,
\label{sthree3}
\end{eqnarray}
where $t_D\sim \omega_G^{-1}$ is the dynamical time. Collective effects tend to reduce the value
of the relaxation time as explained in Appendix \ref{sec_reduction}. A short historic of the early
development of the kinetic theory of 3D stellar systems with many references is given in
\cite{aanew}.

\subsection{2D self-gravitating systems}
\label{sec_sgtwo}

If we neglect strong collisions, collective effects, and make a local approximation, the evolution of the system is described by the Vlasov-Landau equation (II-16). In the thermal bath approach, the diffusion tensor is
given by Eq.  (II-45) with
\begin{equation}
\label{sgtwo1}
{\cal D}_2=\frac{1}{2(2\pi)^{\frac{3}{2}}}\frac{v_m^3}{n}k_J^4\int_{0}^{+\infty}  \frac{dk}{k^2}.
\end{equation}
This integral converges at small scales implying that strong collisions are negligible. However, it diverges rapidly (linearly) at large scales implying that spatial inhomogeneity effects are important. If we introduce a large-scale cut-off $k_{min}=k_J$ at the Jeans length, we obtain
\begin{equation}
\label{sgtwo2}
{\cal D}_2=\frac{1}{2(2\pi)^{\frac{3}{2}}}v_m^3 k_J\frac{1}{\Lambda},
\end{equation}
where $\Lambda= n k_J^{-2}\sim N\gg 1$ represents the number of particles in the system.

Spatial inhomogeneity and collective effects are taken into account in the Lenard-Balescu equation written with angle-action variables \cite{angleaction,kindetail,heyvaerts,newangleaction,aanew}. This equation is rigorously valid for a Newtonian potential in $d=2$ (see Sec. \ref{sec_inf}). If we neglect collective effects, we get the Landau equation written with angle-action variables. These equations converge at large scales since they take into account the finite extent of the system.  If we make a local approximation, and introduce a large scale cut-off at the Jeans length, we get the Vlasov-Landau equation (II-16). Contrary to the 3D case, the local approximation is not very accurate in 2D because the divergence at large scales in the diffusion coefficient (\ref{sgtwo1}) is linear instead of logarithmic. Therefore, the value of the diffusion coefficient is strongly dependent on the large-scale cut-off. However,  the Vlasov-Landau equation (II-16) may provide a reasonable approximation of the true dynamics of self-gravitating systems
in $d=2$ by properly ``adapting'' the value of the large-scale cut-off.

{\it Conclusion:} The evolution of the system as a whole is described  by Lenard-Balescu equation
 (taking into account collective effects) or by the Landau equation (ignoring collective effects)
 written with angle-action variables \cite{angleaction,kindetail,heyvaerts,newangleaction,aanew}. The evolution of a test particle in a thermal bath is described by the Fokker-Planck equation written with angle-action variables. If we neglect collective effects, make a local approximation, and introduce a large-scale cut-off at the Jeans length, the evolution of the system as a whole is described by the Vlasov-Landau
equation (II-16) or (II-17) with $K_2={2\pi G^2 m}/{k_J}$. In the thermal bath approach, we get the Fokker-Planck equation (II-56) with
the diffusion tensor given by Eqs. (II-51) and (II-52) where ${\cal D}_2$ given by Eq. (\ref{sgtwo2})\footnote{In $d=2$, the velocity dispersion of a self-gravitating system in a steady state is exactly given by $\langle v^2\rangle=2v_m^2=GM/2$ (this result can
 be derived from the virial theorem \cite{cras,n2}). Statistical equilibrium states exist at a
 unique temperature $k_B T=GMm/4$. On the other hand, the density profile of a
 distribution of field particles at statistical equilibrium,  and the 
gravitational potential that it generates, are
known analytically (see, e.g., \cite{tc,virial2} and references therein). They are given by 
$\rho(r)=\rho_0/(1+\pi\rho_0r^2/M)^2$ and $\Phi(r)=\frac{GM}{2}\ln(\frac{M}{\pi\rho_0}+r^2)$ where $\rho_0$ is the central density of 
the cluster which parameterizes the series of equilibria. These expressions may be substituted 
in the local diffusion coefficient (\ref{sgtwo2}) and in the advection term of the
Fokker-Planck equation (II-56).}. Using the results of Sec. \ref{sec_powerlaw}, we find that the relaxation time is given by
\begin{eqnarray}
t_R\sim \frac{v_m^2}{G^{3/2}n^{1/2}m^{3/2}}  \sim N t_D,
\label{sgtwo3}
\end{eqnarray}
where $t_D\sim \omega_G^{-1}$ is the dynamical time. Collective effects tend to reduce the value of the relaxation time as explained in Appendix  \ref{sec_reduction}.

In a recent paper, Marcos \cite{marcos} developed a kinetic theory of self-gravitating systems in two dimensions.
He showed through extensive numerical simulations that a local approximation may be
implemented in 2D gravity provided that a large-scale cut-off is introduced (it is a fraction of the
system's size).
He also derived the diffusion and friction coefficients from an extension of
the Chandrasekhar binary collision theory. This leads
to the expressions (II-B6-a) and   (II-B6-b) written in terms of the Rosenbluth potentials (II-B7). He then 
simplified these
expressions for a
Maxwellian distribution leading to his equations (7a) and (7b). We
previously \cite{landaud,paper1} obtained the simpler expressions (II-35), (II-51) and (II-52) from the Landau
approach. As explained in Appendix \ref{sec_lc}, the two approaches (which make a local approximation
and neglect collective effects) are equivalent. The collisional relaxation of 2D self-gravitating systems
has also been studied by Teles {\it et al.} \cite{teles} using  a simplified dynamics.

\subsection{1D self-gravitating systems}
\label{sec_dawg}

If we neglect strong collisions, collective effects, and make a local approximation, the evolution of the system is described by the Vlasov-Landau equation (II-16). In the thermal bath approach, the diffusion coefficient
is given by Eqs. (II-46) and (II-54) leading to
\begin{equation}
\label{dawg1}
D({v})=\frac{1}{(2\pi)^{{1}/{2}}}\frac{v_m^3}{n}k_J^4  e^{-\frac{1}{2}\beta m v^{2}} \int_{0}^{+\infty}  \frac{dk}{k^3}.
\end{equation}
This integral converges at small scales implying that strong collisions are negligible. However, it diverges rapidly (quadratically) at large scales implying that spatial inhomogeneity effects are important. If we introduce a large-scale cut-off $k_{min}=k_J$ at the Jeans length, we obtain
\begin{equation}
\label{dawg2}
D({v})=\frac{k_J}{2(2\pi)^{1/2}(\beta m)^{3/2}\Lambda}e^{-\frac{1}{2}\beta m v^{2}},
\end{equation}
where $\Lambda= n k_J^{-1}\sim N \gg 1$ represents the number of particles in the system.

Spatial inhomogeneity and collective effects are taken into account in the Lenard-Balescu equation written with angle-action variables \cite{angleaction,kindetail,heyvaerts,newangleaction,aanew}. This equation is rigorously valid for a Newtonian potential in $d=1$ (See Sec. \ref{sec_inf}). If we ignore collective effects, we get the Landau equation written with angle-action variables. These equations converge at large scales since they take into account the finite extent of the system. If we neglect collective effects, make a local approximation, and introduce a large scale cut-off at the Jeans length, we get the Vlasov-Landau equation (II-16). However, the Vlasov-Landau equation is expected to display important discrepancies
with the true dynamics for at least two reasons. (i) First of all, it is clear that we cannot make a local approximation to describe the evolution of the system as a whole. Indeed, in $d=1$ the collision term in the Vlasov-Landau equation (II-16) vanishes so this equation reduces to the Vlasov equation. Therefore, it  implies that the relaxation time scales as $N^2 t_D$ (like for 1D plasmas) while numerical simulations \cite{joyce} of 1D self-gravitating systems show that the relaxation time is of order $N t_D$. Actually, when spatial inhomogeneity is properly accounted for,  additional resonances appear in  the Lenard-Balescu equation (or in the Landau equation) written with angle-action variables, and the collision term is non-zero (see discussion in \cite{kindetail}). This explains why the relaxation time scales as $N t_D$ instead of $N^2 t_D$.
 (ii) On the other hand,
in the thermal bath approach, the Vlasov-Landau equation (II-16) leads to the Fokker-Planck equation  (II-57) with the diffusion 
coefficient (\ref{dawg2})\footnote{In $d=1$, the density profile of a
 distribution of field particles at statistical equilibrium,  and the 
gravitational potential that it generates, are
known analytically (see, e.g., \cite{tc} and references therein). They are given by 
$\rho(x)=(GM^2m/4k_B T)\cosh^{-2}(GMmx/2k_BT)$ and $\Phi(x)=(2k_B T/m)\ln\lbrack \cosh(GMmx/2k_B T)\rbrack$. 
These expressions may be substituted 
in the local diffusion coefficient (\ref{dawg2}) and in the advection term of the
Fokker-Planck equation (II-57).}. Contrary to the 3D case, the local approximation is not valid in 1D because the divergence at large scales in the diffusion coefficient (\ref{dawg1}) is quadratic instead of logarithmic. Therefore, the value of the diffusion coefficient is strongly dependent on the large-scale cut-off. Furthermore, we should not give too much credit on the precise expression (\ref{dawg2}) of the diffusion coefficient. We have seen in the case of 1D plasmas that collective effects change the velocity dependence of the diffusion coefficient (for large velocities). This is probably true also for 1D self-gravitating systems. For all these reasons, the local approximation is not a good approximation in $d=1$.

{\it Conclusion:} The evolution of the system as a whole is described by Lenard-Balescu equation 
(taking collective effects into account) or by the Landau equation (neglecting collective effects) 
written with angle-action variables \cite{angleaction,kindetail,heyvaerts,newangleaction,aanew}. The evolution of a test particle in a thermal bath is described by the Fokker-Planck equation written with angle-action variables.    Using the results of Sec. \ref{sec_powerlaw}, we find that the relaxation time  is given by
\begin{eqnarray}
t_R\sim \frac{v_m}{Gm} \sim N t_D,
\label{dawg3}
\end{eqnarray}
where $t_D\sim \omega_G^{-1}$ is the dynamical time. We note that the relaxation time is independent on the density. Collective effects tend to reduce the value of the relaxation time as explained in Appendix  \ref{sec_reduction}. The local approximation cannot be employed for 1D self-gravitating systems.

The collisional relaxation of 1D self-gravitating systems has been studied by Miller \cite{miller},
Valageas \cite{valageas}, Joyce and 
Worrakitpoonpon \cite{joyce} and Sano \cite{sanograv}.

\section{The HMF model}
\label{sec_hmf}

The HMF model \cite{ar} consists in $N$ particles moving on a circle and interacting via a potential $u(\theta-\theta')=\frac{1}{N}\lbrack 1-\epsilon\cos(\theta-\theta')\rbrack$ with $\epsilon=+1$ in the attractive case and $\epsilon=-1$ in the repulsive case. The Fourier transform of the potential is $\hat{u}_n=\frac{1}{2N}(2\delta_{n,0}-\epsilon\delta_{n,\pm 1})$ and its normalized Fourier transform is $\eta_n=\frac{1}{2}\beta(-2\delta_{n,0}+\epsilon\delta_{n,\pm 1})$. The repulsive HMF model does not display any phase transition. Its statistical equilibrium states are always spatially homogeneous (for any energy and temperature). The attractive HMF model displays a second order phase transition. Its statistical equilibrium states are spatially  homogeneous for $E>E_c=3/4$ (i.e. $T>T_c=1/2$) and spatially  inhomogeneous for $E<E_c$ (i.e. $T<T_c$); when $E<E_c$ (i.e. $T<T_c$), the spatially homogeneous phase is unstable \cite{cdr}. Here, we restrict ourselves to the stable homogeneous phase (the kinetic theory of the spatially inhomogeneous HMF model can be treated with angle-action variables 
as in \cite{angleaction,kindetail,heyvaerts,newangleaction,aanew}).

If we neglect collective effects, the relaxation of a test particle in a thermal bath is described by the Fokker-Planck equation (II-42) with a diffusion coefficient given by Eqs. (II-46) and (II-54)  leading to
\begin{equation}
\label{hmf0}
D(v)=\frac{1}{N}\frac{\pi}{2}\left (\frac{\beta}{2\pi}\right )^{1/2}e^{-\frac{1}{2}\beta v^2}.
\end{equation}
If collective effects are taken into account, the diffusion coefficient is given by Eqs. (II-43) and (II-44) leading to
\begin{equation}
\label{hmf1}
D(v)=\frac{1}{N}\frac{\sqrt{2\pi\beta}\, e^{-\frac{1}{2}\beta v^2}}{\left (2\epsilon-\beta+\sqrt{2}\beta^{3/2}v e^{-\frac{1}{2}\beta v^{2}}\int_{0}^{\sqrt{\frac{\beta}{2}}v}e^{y^{2}}dy\right )^{2}+\frac{\pi}{2}\beta^3 v^2  e^{-\beta v^2}}.
\end{equation}
It has the asymptotic behaviors
\begin{equation}
\label{hmf2}
D(0)=\frac{1}{N}\frac{\sqrt{2\pi\beta}}{(2\epsilon-\beta)^{2}},\qquad D(v)\sim_{\pm\infty} \frac{1}{N}\frac{\pi}{2}\left (\frac{\beta}{2\pi}\right )^{1/2}e^{-\frac{1}{2}\beta v^2}.
\end{equation}
The exact expression (\ref{hmf1}) of the diffusion coefficient reduces to the approximate expression (\ref{hmf0}) at high temperatures $T\rightarrow +\infty$ or for large
velocities $|v|\rightarrow +\infty$. For low velocities $v\rightarrow 0$, the diffusion coefficient (\ref{hmf2}-a) can be obtained from Eq. (\ref{hmf1}) by replacing  the ``dressed'' potential ${\hat u}_n^{dressed}=\hat{u}_n/|\epsilon(n,n v)|$ by the Debye-H\"uckel  potential $\hat{u}_n^{DH}=\hat{u}_n/|\epsilon(n,0)|$. In physical space, it reads
\begin{equation}
\label{hmfdressed}
u_{DH}(\theta-\theta')=\frac{1}{N}\left\lbrack 1-\frac{\epsilon}{1-\frac{1}{2}\epsilon\beta}\cos(\theta-\theta')\right\rbrack.
\end{equation}
In the attractive case $\epsilon=+1$, the diffusion coefficient $D(0)$ diverges as $(2-\beta)^{-2}$ at the critical point $\beta_c=2$. Introducing the dynamical time $t_D\sim 2\pi/v_m$, we find that the relaxation time $t_R^{bath}\sim v_m^2/D$ scales like
\begin{equation}
\label{hmf3}
t_R^{bath}\sim \left (2T-\epsilon \right )^2 Nt_D,
\end{equation}
where we have estimated the diffusion coefficient at $v=0$ (i.e. in the Debye-H\"uckel approximation). This formula shows that the relaxation time depends on the temperature. At high temperatures, we can neglect collective effects and we obtain $t_R^{bath}\sim 4T^{2} Nt_D$. On the other hand, if we take collective effects into account, we find that the relaxation time tends to zero at the critical point $T_c=1/2$ in the attractive case $\epsilon=+1$. Therefore, collective effect reduce  the relaxation time close to the critical point. The same conclusion has been reached for self-gravitating systems (see Appendix \ref{sec_reduction}). However, this prediction could be tested
numerically more easily with the HMF model since the potential of interaction is restricted to one Fourier mode and the system is finite.

The evolution of the system as a whole is given by the Lenard-Balescu equation (II-3) which reduces to $\partial f/\partial t=0$ for spatially homogeneous distributions in $d=1$ (see Eq. (II-11)). This implies that the relaxation time of the system as a whole is
\begin{eqnarray}
t_R^{whole}> N t_D.
\label{hmf4}
\end{eqnarray}
If the system remains always spatially homogeneous, we expect that the relaxation time scales as $N^2 t_D$,  like for 1D plasmas (see Sec. \ref{sec_daw}), due to the absence of resonances at the order $1/N$. If the system remains always spatially inhomogeneous, we expect  that the relaxation time scales as $N t_D$, like for 1D stellar systems (see Sec. \ref{sec_dawg}), due to additional resonances at the order $1/N$ brought by spatial inhomogeneity. Finally, if the system is initially spatially homogeneous but, due to the development of correlations (finite $N$ effects), becomes Vlasov unstable and experiences a dynamical phase transition from a homogeneous phase to an inhomogeneous phase \cite{campaall}, we expect that the relaxation time is intermediate between the two previous scalings $Nt_D$ and $N^2t_D$. This argument is consistent with numerical results showing that the relaxation time scales as $N^{1.7}t_D$ in that case \cite{yamaguchi}.

Since the distribution of the field particles does not change on a timescale of the order $Nt_D$ in the homogeneous case, we can consider the relaxation of a test particle in an out-of-equilibrium bath (see Sec. IV. C of Paper II). This relaxation process is described by the Fokker-Planck equation (II-63) with a diffusion coefficient given by Eqs. (II-58) and (II-61). When collective effects are neglected, we obtain
\begin{equation}
\label{fmh6} D(v)=\frac{1}{N}\pi^2  f(v).
\end{equation}
When collective effects are taken into account, we get
\begin{equation}
\label{fmh5} D(v)=\frac{1}{N}\pi^2  \frac{f(v)}{|\epsilon(1,v)|^2}=\frac{1}{N} \frac{\pi^2 f(v)}{\left\lbrack 1+\epsilon\pi {\cal P}\int_{-\infty}^{+\infty}\frac{f'(u)}{u-v}\, du\right\rbrack^2+\pi^4 f'(v)^2}.
\end{equation}
If $f(v)$ decreases sufficiently rapidly, we find that the exact diffusion coefficient (\ref{fmh5}) behaves as $D(v)\sim N^{-1}\pi^2 f(v)$  for $|v|\rightarrow +\infty$ as when collective effects are neglected. For the isothermal distribution (thermal bath), we recover Eq. (\ref{hmf1}). For polytropic (Tsallis) distributions, the diffusion coefficient $D(v)$ has been computed numerically in \cite{cvb}. For the waterbag distribution, using Eq. (II-64), we get
\begin{equation}
\label{fmh7} D(v)=\frac{1}{N} \frac{\pi}{4v_m}\left (\frac{v_m^2-v^2}{v_m^2-\frac{\epsilon}{2}-v^2}\right )^2,
\end{equation}
if $-v_m\le v\le v_m$ and $D(v)=0$ otherwise. As noted in \cite{cvb}, for the attractive HMF model, the diffusion coefficient diverges when the velocity $v$ of the test particle is equal to the pulsation  $\omega=(v_m^2-1/2)^{1/2}$ of the wave arising from the slightly perturbed distribution of the field particles. This divergence occurs for any stable
distribution ($v_m\ge 1/\sqrt{2}$ i.e. $E\ge E_c$) because, for the waterbag distribution, the modes are purely oscillatory. For a single humped distribution with a maximum at $v=v_0$, the diffusion coefficient (\ref{fmh5}) diverges only at the critical point ($E=E_c$) for $v=v_0$. Indeed, the critical point $E=E_c$ is such that $\epsilon(1,v_0)=0$ \cite{nyquist}. For the isothermal distribution, this implies that $D(0)$ diverges when $T\rightarrow T_c$ in agreement with Eq. (\ref{hmf2}-a).

The kinetic theory of the HMF model was first considered by Inagaki \cite{inagakikin}. He tried to adapt the Lenard-Balescu equation to that model but, unfortunately,  his paper contains mistakes that led him to incorrect conclusions. The relaxation of a test particle in a thermal bath was considered by Bouchet \cite{bouchet} who derived the diffusion coefficient (\ref{hmf1}) from the study of the stochastic process of equilibrium fluctuations. In later works by 
Bouchet and Dauxois \cite{bd} and by Chavanis \cite{cvb,hb2}, it was realized that the kinetic theory of the HMF model could be performed by adapting the results of plasma physics to this specific system. Indeed, the previous results can be viewed as particular cases of the general results presented in Paper II.

\section{Conclusion}
\label{sec_conclusion}

In this series of papers, we have completed the literature on the kinetic theory of systems with long-range interactions. A general formalism has been developed in Papers I and II and it has been applied to specific potentials of interaction in the present paper. We have considered pure power-law potentials of interaction and we have explained how the divergences that occur in the Landau equation may be cured by taking into account strong collisions (Boltzmann) or collective effects (Lenard-Balescu) depending on the value of the index $\gamma$. We have also treated the case of plasmas and stellar systems in $d$ dimensions and the case of the HMF model with attractive or repulsive interactions. The relation to previous works has been discussed in detail. An interest of our presentation is to provide a ``unified'' picture of the subject.

\appendix

\section{The different kinetic equations}
\label{sec_bbgky}

The standard kinetic equations (Boltzmann, Fokker-Planck, Vlasov, Landau, Lenard-Balescu) may be derived from the Klimontovich equation by using a quasilinear approximation \cite{pitaevskii} as discussed in Paper I. They may equivalently be derived from the Liouville equation by neglecting three-body correlations in the BBGKY hierarchy \cite{ichimaru,nicholson,balescubook}. In this Appendix, writing the BBGKY hierarchy in a symbolic form, we show the connection between these different equations and discuss their domains of validity without going into technical details.

For spatially homogeneous systems, the first two equations of the BBGKY hierarchy may be  written symbolically as
\begin{eqnarray}
\label{bbgky0}
\frac{\partial f}{\partial t}=C[g],\qquad \frac{\partial g}{\partial t}+({\cal L}_0+{\cal L}')g+{\cal C}[f,g]+{\cal T}[h]={\cal S}[f],
\end{eqnarray}
where $f$ is the one-body distribution function, $g$ the two-body correlation function, and $h$ the three-body correlation function. In the first equation, the collision term $C[g]$ describes the effect of two-body correlations on the evolution of the distribution function. In the second equation, ${\cal L}={\cal L}_0+{\cal L}'$ is a two-body Liouvillian operator where ${\cal L}_0$ describes the free motion of the particles and ${\cal L}'$ describes the exact two-body interaction.
 The term ${\cal C}[f,g]$ describes collective effects and the term ${\cal T}[h]$ describes three-body correlations. Finally, ${\cal S}[f]$ is a source term depending on the one-body distribution function.

For systems with short-range interactions, the potential decreases at large distances as $r^{-\gamma}$ with $\gamma>d$ (i.e. $\alpha>2$). In general, there is also a repulsion at short distances like in the Lennard-Jones potential or in the hard sphere potential. Therefore, the potential has a finite range $l_0$. In the dilute limit $n l_0^d\ll 1$ (this corresponds to $l_0\ll l$), we can neglect collective effects and three-body collisions (${\cal C}={\cal T}=0$). Therefore, the evolution of the system is driven by weak and strong collisions. The coupled equations (\ref{bbgky0}) with ${\cal C}={\cal T}=0$ lead to the
Boltzmann equation \cite{balescubook}.

For systems with long-range interactions, the potential decreases at large distances as $r^{-\gamma}$ with $\gamma<d$ (i.e. $\alpha<2$). In the the weak coupling approximation $g\ll 1$ or $\Lambda=n\lambda_D^d\gg 1$ (this corresponds to $\lambda_D\gg l$), at the order $1/\Lambda$, we can neglect three-body correlations (${\cal T}=0$) and strong collisions (${\cal L}'=0$). Therefore, we can replace the exact two-body interaction of the particles by their free motion (${\cal L}_0+{\cal L}'\simeq {\cal L}_0$) which amounts to making a linear trajectory approximation. The evolution of the system is driven by weak collisions and collective effects. The coupled equations (\ref{bbgky0}) with ${\cal L}'={\cal T}=0$ lead to the Lenard-Balescu equation \cite{balescubook}. If, in addition, we neglect collective effects (${\cal C}=0$), we get the Landau equation\footnote{We must be careful, however, that for potentials that are singular at $r=0$, strong collisions may be important at small scales (in that case, the expansion of the BBGKY hierarchy in powers of $g=1/\Lambda$ is not uniformly convergent). This may invalidate the Lenard-Balescu and Landau equations. This difficulty has been discussed in Sec. \ref{sec_powerlaw}
for purely power-law potentials.}.

The Landau equation (${\cal L}'={\cal T}={\cal C}=0$) is intermediate between the Boltzmann and the Lenard-Balescu equation. It describes the effect of weak collisions but ignores strong collisions and collective effects. It can be obtained from the Lenard-Balescu equation (${\cal L}'={\cal T}=0$) by neglecting collective effects (${\cal C}=0$) or from the Boltzmann equation (${\cal C}={\cal T}=0$) by considering the limit of small deflexions (${\cal L}'=0$).

Actually, these kinetic equations describe the effect of collisions at different scales (the scale $\lambda$ may be interpreted as the impact parameter). For $\lambda\sim \lambda_L$ (small impact parameters), the collisions are strong and we must solve the two-body problem exactly and take into account the bending of the trajectories of the particles. For $\lambda\sim l$ (intermediate impact parameters), the collisions are weak and we can make a weak coupling, or straight line, approximation. For $\lambda\sim \lambda_D$ (large impact parameters), we must take collective effects into account. The Boltzmann equation is valid for $\lambda\ll \lambda_D$. It describes strong collisions ($\lambda\sim\lambda_L$) and weak collisions ($\lambda\sim l$). The Lenard-Balescu equation is valid for $\lambda\gg \lambda_L$. It describes weak  collisions ($\lambda\sim l$) and collective effects ($\lambda\sim \lambda_D$). The Landau equation is valid for $\lambda_L\ll \lambda\ll \lambda_D$. It describes  weak collisions. When we go beyond the domains of validity of these equations, divergences occur, and appropriate cut-offs must be introduced.

For spatially inhomogeneous systems (see \cite{aanew} for more details), the first two equations of the BBGKY hierarchy may be  written symbolically as
\begin{eqnarray}
\label{bbgky1}
\frac{\partial f}{\partial t}+({\cal V}_0+{\cal V}_{m.f.}[f]) f=C[g],\qquad \frac{\partial g}{\partial t}+({\cal L}_0+{\cal L}'+ {\cal L}_{m.f.}[f])g+{\cal C}[f,g]+{\cal T}[h]={\cal S}[f].
\end{eqnarray}
In the first equation, ${\cal V}={\cal V}_0+{\cal V}_{m.f.}$ is the Vlasov operator taking into account the free motion ${\cal V}_0$ of the particles and the advection by the mean field ${\cal V}_{m.f.}$.
In the second equation, the term  ${\cal L}_{m.f.}$ in the two-body Liouvillian takes into account the effect of the mean field in the two-body problem.

For systems with long-range interactions, in the weak coupling approximation $g\ll 1$ or $\Lambda=n\lambda_J^d\gg 1$ (this corresponds to $\lambda_J\gg l$),  at the order $1/\Lambda$, we can neglect three-body correlations (${\cal T}=0$) and strong collisions (${\cal L}'=0$). Therefore, we can replace the exact two-body dynamics of the particles by their mean field motion (${\cal L}_0+{\cal L}'+{\cal L}_{m.f.}\simeq {\cal L}_0+{\cal L}_{m.f.}$). The evolution of the system is driven by weak collisions, collective effects, and spatial inhomogeneity. The coupled equations (\ref{bbgky1}) with ${\cal L}'={\cal T}=0$ lead to the Lenard-Balescu equation written with angle-action variables \cite{angleaction,kindetail,heyvaerts,newangleaction,aanew}\footnote{We can use angle-action variables because for $\Lambda\gg 1$ the collisional evolution is a slow process which allows us to make an adiabatic approximation.}. If, in addition, we neglect collective effects (${\cal C}=0$), we get the Landau equation  written with angle-action variables. Finally, if we neglect collisions (which is rigorously valid for $\Lambda\rightarrow +\infty$) we get the Vlasov equation.

\section{The modified Landau equation}
\label{sec_landaumodif}

For pure power-law potentials with $\alpha>\alpha_c$ (i.e. $\gamma>\gamma_c$ or $d>3$ for the Coulombian or Newtonian potential), the diffusion coefficient (\ref{pl2}) is divergent at small scales ($k\rightarrow +\infty$) while it is convergent at large scales ($k\rightarrow 0$). In that case, it is necessary to take strong collisions  into account, while collective effects (and spatial inhomogeneity effects for attractive interactions) may be neglected (see Sec. \ref{sec_mls}). The dynamical evolution of the system is then described by the Boltzmann equation. However, for long-range potentials ($\alpha<2$, i.e. $\gamma<d$), weak collisions dominate over strong collisions when $\Lambda\gg 1$. We can therefore expand the Boltzmann equation for weak deflexions $|\Delta {\bf v}|\ll 1$. This is equivalent to starting from the Fokker-Planck equation (II-25) and computing the first and second moments of the velocity increments by a binary encounter theory. However, in the calculation of $\langle \Delta {v}_i \rangle$ and $\langle \Delta v_i \Delta v_j\rangle$, we must use the exact trajectory of the particles (and take into account the strong deflexions for collisions with small impact parameters) in order to have convergent expressions. This leads to the modified Landau equation, written below in $d=3$ for $\alpha>\alpha_c=0$ (i.e. $\gamma>\gamma_c=1$):
\begin{equation}
\frac{\partial f}{\partial t}=\frac{A}{m}\left (\frac{e^2}{m}\right )^{2/(1+\alpha)}\frac{\partial}{\partial v_i}\int d{\bf v}' \frac{w^2\delta_{ij}-w_iw_j}{w^{(3-\alpha)/(1+\alpha)}}\left (\frac{\partial}{\partial {v}_{j}}-\frac{\partial}{\partial {v'}_{j}}\right )f({\bf v},t)f({\bf v}',t),
\label{landaumodif1}
\end{equation}
where  $A$ is a coefficient (depending on $\alpha$) related to the integration over the impact parameter. For $\alpha>\alpha_c$, the modified Landau equation (\ref{landaumodif1}) does not present
any divergence (see Sec. \ref{sec_mls}). For $\alpha<\alpha_c$, the modified Landau equation (\ref{landaumodif1}) presents an algebraic divergence at large scales and it is not valid anymore. In that case, it must be replaced by the Lenard-Balescu equation (II-3) as explained in Sec. \ref{sec_inf}. For $\alpha=\alpha_c$, Eq. (\ref{landaumodif1}) reduces to the ordinary Landau equation (II-9). This equation presents a logarithmic divergence at large scales which can be cured by introducing a large-scale cut-off at the Debye length $\lambda_{max}=\lambda_D$ (properly justified) as explained in Sec. \ref{sec_eq}. On the other hand, Eq. (\ref{landaumodif1}) does {\it not}  diverge at small scales since the effect of strong collisions is taken into account. In that case, the Landau length appears naturally in the diffusion coefficient $K_3=(2\pi e^4/m^3)\ln(\lambda_{max}/\lambda_L)$ (see the discussion in Appendix \ref{sec_lc}). We emphasize that, for $\alpha>\alpha_c$, the modified Landau equation (\ref{landaumodif1}) differs from the ordinary Landau equation (II-9) since the power of the relative velocity $w$ appearing in the denominator of the collision term is $(3-\alpha)/(1+\alpha)$ 
instead of $3$ (in $d=3$ dimensions). From Eq. (\ref{landaumodif1}) we find that the relaxation time scales as
\begin{equation}
t_R\sim \frac{v_m^{\frac{3-\alpha}{1+\alpha}}}{\left (\frac{e^2}{m}\right )^{\frac{2}{1+\alpha}}n}
\sim \Lambda^{\frac{1-\alpha}{1+\alpha}}t_D,
\label{landaumodif2}
\end{equation}
in agreement with Eq. (\ref{pl5}). 

The modified Landau equation (\ref{landaumodif1}) was introduced by Potapenko {\it et al.} \cite{potapenko}. The Coulombian case corresponds to $\gamma=1$ and the case of Maxwell molecules corresponds to $\gamma=4$. The potential is soft for $1\le \gamma< 4$ and hard for $\gamma>4$. Some applications will be discussed elsewhere.

\section{Reduction of the relaxation time due to collective effects for attractive interactions}
\label{sec_reduction}

For attractive power-law potentials of interaction with $\alpha\le \alpha_c$, the evolution of the system is described by the Lenard-Balescu equation written with angle-action variables \cite{angleaction,kindetail,heyvaerts,newangleaction,aanew}. This equation takes  spatial inhomogeneity and collective effects into account. Unfortunately, it is extremely complicated. If we make a local approximation, we obtain the Vlasov-Lenard-Balescu equation (II-15). However, this equation presents a strong divergence at the Jeans scale. Indeed, in the thermal bath approximation, the diffusion tensor  is given by Eq. (II-39) with
\begin{equation}
\label{pl7b}
{\cal D}_d(x)=\frac{1}{2(2\pi)^{d-\frac{1}{2}}}\frac{v_m^3 k_J}{\Lambda}\int_0^{+\infty}  \frac{\kappa^d}{\left\lbrack \kappa^{2-\alpha}-B(x)\right\rbrack^{2}+C(x)^{2}}\, d{\kappa}.
\end{equation}
If we make the Debye-H\"uckel approximation, or consider small velocities $|{\bf v}|\rightarrow 0$, the diffusion tensor  is given by Eq. (II-45) with
\begin{equation}
\label{pl9}
{\cal D}^{DH}_d=\frac{1}{2(2\pi)^{d-\frac{1}{2}}}\frac{v_m^3 k_J}{\Lambda}\int_0^{+\infty}  \frac{\kappa^d}{(\kappa^{2-\alpha}-1)^{2}}\, d{\kappa},\qquad  t_{R}^{bath}\sim \frac{\Lambda}{\int_0^{+\infty}  \frac{\kappa^d\, d{\kappa}}{(\kappa^{2-\alpha}-1)^{2}}}t_D. \qquad
\end{equation}
Clearly, the integral diverges at $k=k_J$ (i.e. $\kappa=1$). This is of course related to the Jeans instability for a spatially homogeneous system. This divergence does not occur when spatial inhomogeneity is properly accounted for \cite{angleaction,kindetail,heyvaerts,newangleaction,aanew}. This divergence suggests, at a heuristic level, that collective effects (which account for anti-shielding) tend to increase the diffusion coefficient, hence to decrease the relaxation time. This reduction should be particularly strong for a system close to instability due to an enhancement of fluctuations \cite{monaghan,hb5}. These arguments are further developed in \cite{gilbert,weinberg} and in the Appendix E of \cite{aanew}.

These results show that it is not possible to make a local approximation and simultaneously take collective effects into account. The usual procedure is to ignore collective effects, make a local approximation, and introduce a large-scale cut-off at the Jeans length. This is the procedure that is usually followed in stellar dynamics \cite{spitzerbook,bt,hut}. A more rigorous approach is to use the Lenard-Balescu or Landau equations written with angle-action variables \cite{angleaction,kindetail,heyvaerts,newangleaction,aanew}.

\section{Connection between the Landau and the Chandrasekhar kinetic theories}
\label{sec_lc}

In this Appendix, we discuss the relation between the Landau kinetic theory and the Chandrasekhar 
kinetic theory. These kinetic theories were developed independently and their connection was not fully 
realized by early workers on the subject (see footnote 5 in Paper II). Even in modern textbooks of astrophysics \cite{spitzerbook,bt}, the kinetic theory of stellar systems is presented with the approach of Chandrasekhar and the Landau equation is not mentioned.

Landau \cite{landau} developed a kinetic theory of 3D Coulombian plasmas by starting from the Boltzmann 
equation and using a weak deflexion approximation. In his calculations, he approximated the trajectories 
of the particles by straight lines even for collisions with small impact parameters\footnote{This is 
equivalent to starting from the Fokker-Planck equation (II-25)-(II-26) and calculating the 
first and second moments of the velocity increments resulting from a succession of binary 
collisions by using a straight line approximation. This is also equivalent to starting from 
the BBGKY hierarchy and neglecting three-body collisions, collective effects, and 
strong collisions (${\cal T}={\cal C}={\cal L}'=0$) as discussed in Appendix \ref{sec_bbgky}.}. This 
leads to the kinetic equation (II-9) with
\begin{eqnarray}
\label{rosen16}
K_3^{Landau}=\frac{2\pi e^4}{m^3}\ln\left (\frac{\lambda_{max}}{\lambda_{min}}\right ).
\end{eqnarray}
This factor presents a logarithmic divergence at small and large scales that Landau regularized heuristically by introducing a small-scale cut-off at the Landau length ($\lambda_{min}=\lambda_L$) and a large-scale cut-off at the Debye length ($\lambda_{max}=\lambda_D$).

Chandrasekhar \cite{chandra} developed a kinetic theory of 3D stellar systems by starting from the 
Fokker-Planck equation (II-25)-(II-26) and calculating the first and second moments of the velocity 
increments resulting from a succession of binary collisions. In the calculation of 
$\langle \Delta {v}_i \rangle$ and $\langle \Delta v_i \Delta v_j\rangle$, he used the {\it exact} 
trajectory of the stars (i.e. he solved the two-body problem exactly) and took into account 
the strong deflexions due to collisions with small impact parameters\footnote{This is equivalent 
to expanding the Boltzmann equation for weak deflexions while taking the effect 
of strong collisions (i.e. the bending of the trajectories) into account. This is also equivalent to starting from the BBGKY hierarchy and neglecting 
three-body collisions and collective effects (${\cal T}={\cal C}=0$), but retaining 
strong collisions (${\cal L}'\neq 0$),  as discussed in Appendices \ref{sec_bbgky} 
and \ref{sec_landaumodif}.}. In the dominant approximation $\ln N\gg 1$, his
approach completed by Rosenbluth {\it et al.} \cite{rosen} leads to the expressions (II-B6) of 
the diffusion tensor and friction force expressed in terms of the Rosenbluth potentials (II-B7) with
\begin{eqnarray}
\label{rosen17}
K_3^{Chandra}=2\pi m G^2\ln\left (\frac{\lambda_{max}v_m^2}{Gm}\right ).
\end{eqnarray}
This factor presents a logarithmic divergence at large scales that can be regularized heuristically by
 introducing a cut-off at the Jeans length\footnote{Actually, Chandrasekhar introduced  a large-scale
 cut-off at the inter-particle distance $l$ but this choice was later on criticized by many authors.
It is now acknowledged that the proper large-scale cut-off in a stellar system is the Jeans length
$\lambda_J$ which is the gravitational analogue of the Debye length $\lambda_D$ in plasma physics.}.
However, contrary to the Landau approach, this factor does {\it not} present a  logarithmic divergence
 at small scales since the effect of strong collisions is taken into account explicitly in the
Chandrasekhar approach\footnote{The approach of Chandrasekhar takes into account strong collisions
with an impact parameter smaller than the Landau length $\lambda_L$ which yield a deflection at an
angle larger than $90^{o}$. As we have seen, this can suppress the divergence at small scales.
However, the Chandrasekhar approach does not take into account the possibility of forming binary
stars which correspond to bound states with strong correlation between particles (see Sec. V of Paper II).
 In other words, Chandrasekhar only considers hyperbolic trajectories and not elliptical ones
in the two-body problem. The effect of binary stars in the kinetic theory requires a special
 treatment \cite{heggie}.}. As a result, the gravitational Landau length
$\lambda_L=Gm/v_m^2$ appears naturally in the calculations of Chandrasekhar.  Of course, these results
can be translated to the plasma case by replacing $G$ by $(e/m)^2$ and $\lambda_J$ by $\lambda_D$.

The Landau and the Chandrasekhar kinetic theories make the same assumption:
binary encounters and an expansion in powers of the momentum transfer. They
actually differ in the order in which these are introduced. Landau starts from the Boltzmann equation and considers a weak deflexion approximation while Chandrasekhar directly starts from the Fokker-Planck equation but calculates
the coefficients of diffusion and friction with the binary-collision picture. At that level, their theories are equivalent since the Fokker-Planck equation can precisely be obtained from the Boltzmann equation in the
limit of weak deflexions. The crucial difference is that Landau makes a weak coupling assumption and
 ignores strong collisions while Chandrasekhar takes them into account.

Except for this important difference, the calculations of Appendix B of Paper II show that the
 kinetic equation  (II-9) derived by Landau \cite{landau} is equivalent to the kinetic equation (II-B8)
 derived by Chandrasekhar \cite{chandra} and Rosenbluth {\it et al.} \cite{rosen} although they appear
under a different form. This equivalence is not always
apparent in the astrophysical literature (see the discussion in \cite{aanew}). In the astrophysical
 literature, the diffusion and friction coefficients are usually derived from the Chandrasekhar binary
encounter theory leading directly to Eqs. (II-B6)-(II-B7) while they can be obtained
equivalently from the Landau equation (II-B1) leading to Eqs. (II-B3)-(II-B4)
which can be transformed into Eqs. (II-B6)-(II-B7) as we have shown in Appendix B of Paper II.

We stress that the equivalence between the Chandrasekhar approach and the Landau approach is true
 for any dimension $d>1$ and for any long-range potential of interaction (their limitation 
has been discussed in this paper). Therefore, the kinetic theory developed by Gabrielli {\it et al.} \cite{gjm} and Marcos \cite{marcos} based on the Chandrasekhar approach is equivalent to the kinetic theory that we have presented in \cite{landaud} based on the Landau approach. Indeed, essentially the same assumptions are made: spatial homogeneity (or local approximation), neglect of collective effects, and weak coupling approximation. The Lenard-Balescu approach that we have presented in \cite{paper1} is more general since it can take collective effects into account. Finally, the generalized Landau approach presented in \cite{kindetail,aanew} can take into
account spatial inhomogeneity and the generalized  Lenard-Balescu approach presented in \cite{heyvaerts,newangleaction} can take into account spatial inhomogeneity and collective effects.

\section{The choice of the large-scale cut-off}
\label{sec_lsco}

The kinetic theories developed by Landau \cite{landau} in plasma physics and by Chandrasekhar \cite{chandra} in stellar dynamics, modeling the collisional process by a succession of independent weak binary encounters, lead to a diffusion coefficient and a friction force that diverge logarithmically at large impact parameters. These authors heuristically circumvented this problem by introducing a large scale cut-off. 

In the case of plasmas,  Persico \cite{persico}, Landau \cite{landau}, Bohm and Aller \cite{ba},  Prigogine and Balescu \cite{priba},  and Cohen {\it et al.} \cite{cohen} argued that the logarithmic divergence should be cut-off at the Debye length $\lambda_D$. This leads to a Coulombian factor
\begin{equation}
\ln\left (\frac{\lambda_{max}}{\lambda_L}\right )=\ln\left (\frac{\lambda_{D}}{\lambda_L}\right )=\ln\Lambda,
\end{equation}
where $\Lambda=n\lambda_D^3$ is the number of electrons in the Debye sphere. The kinetic theory of Lenard and Balescu proved that the Debye length indeed represents the relevant large-scale cut-off to introduce in the Landau equation (this is exact in the dominant approximation).

In stellar dynamics, the choice of the large-scale cut-off $\lambda_{max}$ created some debate in the early literature. Jeans \cite{jeansbook}, Spitzer \cite{spitzer}, 
Chandrasekhar and von Neumann \cite{cvn}, and  Prigogine and Balescu \cite{priba}  argued that the logarithmic divergence has to be cut-off at the interparticle distance $l\sim n^{-1/3}$ (see Sec. \ref{sec_lengths}). This leads to a Coulombian factor
\begin{equation}
\ln\left (\frac{\lambda_{max}}{\lambda_L}\right )=\ln\left (\frac{l}{\lambda_L}\right )=\frac{2}{3}\ln \Lambda,
\end{equation}
where $\Lambda=n\lambda_J^3\sim N$ is the number of stars in the Jeans sphere. However, Cohen {\it et al.} \cite{cohen} argued that the divergence has to be cut-off at the Jeans scale $\lambda_J$ (of the order of the system's size) which is the gravitational analogue of the Debye length. This leads to a Coulombian factor
\begin{equation}
\ln\left (\frac{\lambda_{max}}{\lambda_L}\right )=\ln\left (\frac{\lambda_J}{\lambda_L}\right )=\ln \Lambda.
\end{equation}
The corresponding relaxation time is lower than the original Chandrasekhar relaxation time by a factor $3/2$. H\'enon \cite{henon}, 
Prigogine and Severne \cite{ps}  argued that if the system were spatially homogeneous and infinite, the relevant large-scale cut-off would be the mean free path $\lambda$ (see Sec. \ref{sec_qss}). This leads to a Coulombian factor
\begin{equation}
\ln\left (\frac{\lambda_{max}}{\lambda_L}\right )=\ln\left (\frac{\lambda}{\lambda_L}\right )=2\ln \Lambda.
\end{equation}
The corresponding relaxation time is lower than the original Chandrasekhar relaxation time by a factor $3$.  Finally, 
Severne and Haggerty \cite{sh} showed that when the spatial inhomogeneity and the finite extent of the system are properly taken into account, there is no large-scale divergence anymore. This suggests that the Jeans scale is the most relevant large-scale cut-off to consider.

\end{document}